\documentclass[preprint,3p,sort&compress]{elsarticle}
\usepackage[intlimits]{amsmath}
\usepackage{amssymb}
\usepackage{graphicx,pstricks}
\usepackage{booktabs}
\usepackage{mathrsfs,slashed}
\usepackage{hyperref}
\usepackage{epsfig}
\usepackage{color}

\newcommand{\bea}{\begin{eqnarray}}
\newcommand{\eea}{\end{eqnarray}}

\newcommand{\nn}{\nonumber}

\newcommand{\viz}{{\it viz., }}
\newcommand{\ie}{{\it i.e., }}
\newcommand{\eg}{{\it e.g., }}
\newcommand{\ii}{{\rm i}}
\newcommand{\ee}{\,{\rm e}}

\DeclareMathOperator{\RE}{Re}
\DeclareMathOperator{\IM}{Im}

\DeclareMathOperator{\Tr}{Tr} 
\DeclareMathOperator{\tr}{tr}

\begin{document}
\title{Generalized Beth--Uhlenbeck approach to mesons and diquarks\\
in hot, dense quark matter
}
\author[wroclaw,dubna]{D.~Blaschke}
\ead{blaschke@ift.uni.wroc.pl}

\author[darmstadt]{M.~Buballa}
\ead{michael.buballa@physik.tu-darmstadt.de}

\author[wroclaw]{A.~Dubinin}
\ead{aleksandr.dubinin@ift.uni.wroc.pl}

\author[rostock]{G.~R\"opke}
\ead{gerd.roepke@uni-rostock.de}

\author[wroclaw,darmstadt]{D.~Zablocki}
\ead{zablocki@ift.uni.wroc.pl}

\address[wroclaw]{
	Institute for Theoretical Physics,
	University of Wroc{\l}aw,
	pl. M. Borna 9, 
	50-204 Wroc{\l}aw, Poland
}
\address[dubna]{
	Laboratory for Theoretical Physics,
	Joint Institute for Nuclear Research,
	Joliot-Curie ul. 6,
	141980 Dubna, Russia
}
\address[darmstadt]{
	Institute for Nuclear Physics,
	Technical University of Darmstadt,
	Schlossgartenstr. 2,
	64289 Darmstadt, Germany
}
\address[rostock]{
	Institut f\"ur Physik,
	Universit\"at Rostock,
	Universit\"atsplatz 3,
	18055 Rostock, Germany
}

\date{\today}
\begin{abstract}
An important first step in the program of hadronization of chiral quark models 
is the bosonization in meson and diquark channels. 
This procedure is presented at finite temperatures and chemical potentials 
for the SU(2) flavor case of the NJL model
with special emphasis on the mixing between scalar meson and scalar diquark
modes which occurs in the 2SC color superconducting phase.
The thermodynamic potential is obtained in the gaussian approximation for the 
meson and diquark fields and it is given the Beth-Uhlenbeck form. 
This allows a detailed discussion of bound state dissociation in hot, dense 
matter (Mott effect) in terms of the in-medium scattering phase shift of 
two-particle correlations. 
It is shown for the case without meson-diquark mixing that the phase shift 
can be separated into a continuum and a resonance part. 
In the latter, the Mott transition manifests itself by a change of the 
phase shift at threshold by $\pi$ in accordance with Levinson's theorem, when 
a bound state transforms to a resonance in the scattering continuum.
The consequences for the contribution of pionic correlations to the pressure 
are discussed by evaluating the Beth-Uhlenbeck equation of state in different
approximations.  
A similar discussion is performed for the scalar diquark channel in the 
normal phase.
Further developments and applications of the developed approach are outlined.
\end{abstract}
\maketitle

\section{Introduction}

One of the long-standing problems of low-energy hadron physics concerns the 
hadronization of QCD, i.e., the transformation of QCD formulated as a gauge 
theory of quark and gluon fields to an effective theory formulated in terms of
hadrons, the observable low-energy degrees of freedom.
The general idea is to start from the path integral representation of the 
partition function (generating functional) of QCD and to ``integrate out'' the 
elementary quark and gluon degrees of freedom, leaving an effective theory in 
terms of collective hadronic degrees of freedom. 
This task may be schematically depicted as the map
\begin{equation}
\label{program}
\mathscr{Z}_{\rm QCD} = \int \mathscr{D}\bar{q} \mathscr{D}{q} \mathscr{D}A
\exp\{\mathscr{A}_{\rm QCD}[\bar{q}, q, A]\} 
\longrightarrow \int \mathscr{D}M \mathscr{D}\bar{B}\mathscr{D}B
\exp\{\mathscr{A}_{\rm eff}[\bar{B}, B, M]\} ~,
\end{equation}
where $\mathscr{A}_{\rm QCD}[\bar{q}, q, A]$ is the Euclidean action functional
of QCD in terms of quark $(\bar{q}, q)$ and gluon $(A)$ fields, while 
$\mathscr{A}_{\rm eff}[\bar{B}, B, M]$ is an effective low-energy action 
functional in terms of baryon $(\bar{B}, B)$ and meson $(M)$ fields.
Since the map (\ref{program}) involves, among others, the still open problem 
of quark confinement, we do not aim at a solution on the level of mathematical 
rigor. 
A slightly modified, solvable version of this program (\ref{program}) does 
start from the global color model of QCD where all nonabelian aspects of the 
gluon sector are absorbed into nonperturbative gluon n-point functions 
(Schwinger functions) starting with the gluon propagator (2-point function) 
coupling to quark currents via nonperturbative vertex functions (3-point 
functions),
see \cite{Roberts:1994dr,Tandy:1997qf,Alkofer:2000wg,Roberts:2000aa} 
for basic reviews. 

After Fierz rearrangement of the current-current interaction 
\cite{Kleinert:1976xz,Ebert:1976rh}, the theory obtains the Yukawa-like form 
with Dirac quark fields coupled to the spectrum of meson and diquark channels,
thus generating 4-fermion interactions in the relevant quark-antiquark and 
quark-quark bilinears  \cite{Cahill:1988zi,Reinhardt:1989rw}, see also 
\cite{Ebert:1994mf}.

The hadronization proceeds further via the two-step Hubbard-Stratonovich
procedure by: 
1) introducing auxiliary collective fields and functional integration over 
them such that the 4-fermion interaction terms are replaced by Yukawa 
couplings to the collective fields, and 
2) integrating out the quark fields so that an effective theory results which 
is formulated in meson and diquark degrees of freedom, highly nonlinear due to 
the Fermion determinant, recognizable as ``$\Tr\ln$'' terms in the effective 
bosonized action.  

In this way one obtains a bosonized chiral quark model, but not yet the desired
hadronization in terms of mesons and baryons, the physical degrees of freedom 
of low-energy QCD and hadronic matter.  
A possible scheme for introducing baryons as quark-diquark bound states and 
integrating out the colored and therefore not asymptotically observable 
diquark fields has been suggested by Cahill and collaborators 
\cite{Cahill:1988zi,Cahill:1988bh,Burden:1988dt,Praschifka:1986nf}.
It was afterwards elaborated by Reinhardt \cite{Reinhardt:1989rw} and 
developed further by including the solitonic aspects of a field theoretic 
description of the nucleon \cite{Zuckert:1996nu}, see also 
Refs.~\cite{Ebert:1985kz,Ebert:1994mf,Christov:1995vm,Golli:1998rf}. 
Introductory reviews on the latter can be found, \eg 
in \cite{Alkofer:1994ph,Alkofer:1995mv,Ripka:1997zb}.

It is the aim of our study to extend the hadronization of effective chiral 
quark models for QCD 
to nonzero temperatures and densities. 
The goal is to investigate the modification of thermodynamic
properties of hadronic matter evoked under extreme conditions by the onset
of (partial) chiral symmetry restoration.
In this first part of our work we investigate in detail the two-particle 
correlations (mesons and diquarks) in a hot and dense quark matter medium 
and the response of their spectral properties to medium dependent mean fields
signaling chiral symmetry breaking (quark mass gap) and color 
superconductivity (diquark pairing gap).
Of central interest is the relation of the meson and diquark spectrum to the 
density and temperature dependence of these order parameters 
(quark mass and pairing gap), the question under which conditions mesons and
diquarks may exist as real bound states or appear only as a correlation in 
the continuum of quark-antiquark and quark-quark scattering states, resp.
The transition of a correlation from the discrete bound state spectrum
to a resonant continuum state is called Mott effect and will be a central
theme for our study.
When it occurs at low temperatures and high densities under fulfillment of
the conditions for Bose condensation, the Mott effect serves as a mechanism
for the BEC-BCS crossover \cite{Sun:2007fc}
which recently became very topical when in atomic
traps this transition could be studied in detail under laboratory conditions.

While numerous works have recently studied the thermodynamics of quark matter
on the mean-field level including the effects of the medium dependence of the 
order parameters, not so much is known beyond the mean field, about hadronic 
correlations and their backreaction to the structure of the model QCD phase
diagram and its thermodynamics.
Here we will elaborate on the generalized Beth-Uhlenbeck form of the equation
of state (EoS) which is systematically extended from studying mesonic 
correlations \cite{Hufner:1994ma} to the inclusion of diquark degrees of 
freedom. 
To that end we will employ a Nambu--Jona-Lasinio-type quark model
with fourpoint interactions in mesonic (quark-antiquark)  and diquark 
(quark-quark) channels.
We shall discuss here the importance of the interplay of the resonant 
states with the residual non-resonant  in the continuum of 
scattering states. Due to the Levinson theorem both contributions have the
tendency to compensate each other in quark matter above the Mott transition
\cite{Wergieluk:2012gd,Yamazaki:2012ux,Dubinin:2013yga}, 
see also \cite{Rossner:2007ik}.
 
The most intriguing questions will occur when on the basis of this in-medium 
bosonized effective chiral quark model the next step of the hadronization 
program nucleon will be performed and diquarks will be ``integrated out'' in
favor of baryons so that nuclear matter can be described in the model
QCD phase diagram.
Exploratory studies within the framework of an effective local NJL-type model
for the quark-diquark interaction vertex have revealed a first glimpse at the
modification of the nucleon spectral function in the different regions of the 
model QCD phase diagram, including chirally restored and color superconducting
phases \cite{Wang:2010iu}. 
 
We will prepare the ground for a Beth-Uhlenbeck description of nuclear matter,
to be discussed in future work. 
In particular, at zero temperature the structure of a Walecka model for 
nuclear matter shall emerge under specified conditions.
Earlier work in this direction \cite{Bentz:2001vc,Bentz:2002um,Lawley:2006ps}
has demonstrated this possibility although 
no unified description of the nuclear-to-quark matter transition has been 
possible and the elucidation of physical mechanisms for the very transition 
between the hadronic and the quark matter phases of low energy QCD has been 
spared out. 
Our study aims at indicating directions for filling this gap 
by providing a detailed discussion of the Mott mechanism for the dissociation 
of hadronic bound states of quarks within the NJL model description 
of low-energy QCD on the example ot two-particle correlations (mesons and 
diquarks).

Our approach can be considered as  complementary
to lattice gauge theory, where $\mathscr{Z}_{\rm QCD} $ is calculated
in ab-initio Monte-Carlo simulations without further approximations other than 
the discretization of space-time~\cite{Bazavov:2011nk,Borsanyi:2010bp}.  
While being more rigorous, lattice calculations are often lacking a clear 
physical interpretation of the results. Moreover, because of the fermion sign
problem, they are restricted to vanishing or small chemical potential. 
In this situation, effective chiral quark model approaches, like the one 
employed here, can give invaluable methodological guidance to elucidate 
how effects like bound state dissociation in hot and dense quark matter,
as seen in spectral functions, may manifest themselves in two-particle 
correlators which are objects accessible to lattice QCD simulations. 

This work is organized as follows. 
In Sect.~\ref{sec:model} we define our model and derive the thermodynamic
potential in mean-field approximation and its corrections to gaussian order
in normal and color su\-per\-conducting matter. 
The corresponding meson and diquark spectra are discussed in 
Sect.~\ref{sec:spectra}.
Sect.~\ref{sec:GBU} is the main part of this article. 
Here we derive a generalized Beth-Uhlenbeck form for the thermodynamic 
potential of two-particle correlations in quark matter as the appropriate 
representation for discussing the 
{\rm Mott} effect, the dissociation of hadronic bound states of quarks 
induced by the lowering of the quark continuum by the chiral symmetry 
restoration at finite temperatures and chemical potentials.  
We evaluate the temperature dependence of the phase shifts for pion, sigma 
meson and scalar diquark interaction channels and obtain the corresponding 
correlation contributions to the EoS.
In closing Sect.~\ref{sec:GBU} we give an outlook to 
further developments and applications of the formalism.
In Sect.~\ref{sec:conclusions} we summarize our main results and present the
conclusions of this work. 

\section{The model: mean field approximation and beyond}
\label{sec:model}

We consider a system of quarks with $N_f=2$ flavor and $N_c=3$ color degrees of
freedom at temperature $T$ and chemical potential $\mu$, described by the 
Nambu--Jona-Lasinio (NJL) type Lagrangian 
\bea
\mathscr{L}=
\mathscr{L}_0+\mathscr{L}_{\rm S}+\mathscr{L}_{\rm V}+\mathscr{L}_{\rm D}~.
\eea
The free part is given by
\bea
\mathscr{L}_{\rm 0}&=&\bar{q}(\ii\slashed\partial-m_0+\mu\gamma_0)q ~,
\eea
with the bare quark mass $m_0$. 
Here we have assumed isospin symmetry, \ie equal masses and chemical potentials
for up and down quarks. 
The quarks interact via local four-point vertices in the scalar-isoscalar, 
pseudoscalar-isovector and vector-isoscalar quark-antiquark channels,
\bea
\mathscr{L}_{\rm S} &=&
G_{\rm S}\left[(\bar{q}q)^2+(\bar{q}\ii\gamma_5\vec\tau q)^2\right]~,\\
\mathscr{L}_{\rm V} &=& - G_{\rm V} (\bar{q}\gamma_\mu q)^2 ~,
\eea
augmented with a quark-quark interaction in the scalar color-antitriplet 
channel,
\bea
\mathscr{L}_{\rm D} &=& G_{\rm D}\sum_{A=2,5,7}
	(\bar{q} \ii\gamma_5\tau_2\lambda_Aq^c)
	(\bar{q}^c\ii\gamma_5\tau_2\lambda_A q) ~.
\eea
Here $q^c= C\bar q^T$ with $C=\ii\gamma^2\gamma^0$ denote the charge conjugate 
quark fields,
$\lambda_A$, $A=2, 5, 7$,  the antisymmetric Gell-Mann matrices in color space 
and $\tau_i$, $i=1,2,3$, the Pauli matrices in flavor space.
$G_{\rm S}$, $G_{\rm V}$ and $G_{\rm D}$ are dimensionful coupling constants. 

Eventually, one should also couple the quarks to an effective Polyakov  
loop variable \cite{Fukushima:2003fw,Megias:2004hj,Ratti:2005jh},  
in order to describe confinement effects in a more realistic way.
Although in principle straightforward, this would lead to the appearance of 
more complicated dispersion relations in the expressions below~\cite{Hansen:2006ee}. 
For simplicity, we therefore leave this extention  for a later publication.

Our model then has five parameters: the bare quark mass $m_0$, 
the coupling constants $G_{\rm S}$, $G_{\rm V}$ and $G_{\rm D}$ and a cutoff
parameter $\Lambda$, which is needed because the interaction is not 
renormalizable.
While $m_0$, $G_{\rm S}$ and $\Lambda$ are typically fitted to vacuum 
properties (mass and decay constant) of the pion and to the chiral condensate
\cite{Klevansky:1992qe,Hatsuda:1994pi,Buballa:2003qv},
the two other coupling constants are less constrained.
They are sometimes related to the scalar coupling via Fierz transformation of 
a color-current interaction, which yields $G_{\rm V} =  G_{\rm S}/2$ and
$G_{\rm D} =  3/4 G_{\rm S}$.
Alternatively, $G_{\rm V}$ can be fixed by fitting the mass of vector mesons, which gives 
higher values~\cite{Vogl:1991qt}.
Another possibility is to constrain $G_{\rm V} $ and $G_{\rm D}$ from compact 
star and heavy-ion phenomenology~\cite{Klahn:2006iw}.
In the present paper, we will specify the parameters only in the context of 
the numerical examples discussed in Sect.~\ref{sec:GBU}, while most of the 
analytic expressions are more general. 
In future extensions of the model one could then try to fix $G_{\rm V}$ and 
$G_{\rm D}$ by fitting baryon and nuclear matter properties.

The bulk thermodynamic properties of the model at temperature $T$ and
chemical potential $\mu$ are encoded in the thermodynamic potential per volume
\bea
	\Omega(T,\mu) = -\frac{T}{V} \ln \mathscr{Z}(T,\mu)
	~,
\eea
with the grand partion function $\mathscr{Z}$ which is given as a functional
integral involving the above Lagrangian, 
\begin{eqnarray}\label{eq:partition-function-1}
	\mathcal{Z}
	=
	\int\mathscr{D}q \mathscr{D}\bar{q}
	\exp\left[ \int{\rm d}^4x_E\,\,{\mathcal L}\right]\,,
\end{eqnarray} 
where $\int{\rm d}^4x_E=\int_0^\beta {\rm d}\tau\int {\rm d}^3{\bf x}$
denotes an integration over the Euclidean four-volume, \ie 
over the three-space with volume $V$ and imaginary time $\tau$ 
restricted to the interval between $0$ and $\beta=1/T$. 
We perform a bosonization by means of Hubbard-Stratonovich transformations.
To this end, we introduce the auxiliary meson fields $\sigma$, $\vec{\pi}$ and 
$\omega_\mu$ in the scalar, pseudoscalar  and vector channel, respectively, as 
well as the complex
auxiliary scalar diquark fields $\Delta_A$ and their complex conjugate fields
$\Delta_A^*$.
Then, after introducing Nambu-Gorkov bispinors 
\begin{eqnarray}
	\Psi
	\equiv
	\frac{1}{\sqrt{2}}
	\left(
	\begin{array}{c}
		q \\q^{c}
	\end{array}
	\right)
	\;\;\;\;\;\;\;
	\bar\Psi
	\equiv
	\frac{1}{\sqrt{2}}
	\left(
	\begin{array}{cc}
		\bar q & \bar{q}^c
	\end{array}
	\right)
\end{eqnarray}
the grand partition function takes the form
\begin{eqnarray}
\label{partition-function-2}
	\mathcal{Z}
	=
	\int\mathscr{D}\sigma \mathscr{D}{\vec \pi}\mathscr{D}\omega_\mu
	\mathscr{D}\Delta_{A}\mathscr{D}\Delta_{A}^{*}
	\exp\left[\int {\rm d}^4 x_E\,
	\left(-\frac{\sigma^2 + {\vec{\pi}}^2}{4 G_S}
	-\frac{\Delta^*_A\Delta_A}{4 G_D}\right)\right]
	\int \mathscr{D}\Psi \mathscr{D}\bar\Psi
	\exp\left[\int {\rm d}^4{\bf x_E}\, 
	\bar\Psi\,S^{-1}\Psi\right]~,
\end{eqnarray}
with the inverse Nambu-Gorkov propagator defined as
\begin{eqnarray}
\label{inverse-propagator}
	S^{-1}
	\equiv
	\left(
	\begin{array}{cc}
		\ii\slashed\partial+\mu\gamma_0 
		+\gamma^\mu \omega_\mu - m_0 -\sigma 
		-\ii\gamma_5{\vec \tau}\cdot{\vec{\pi}}& 
		\ii\Delta_A\gamma_5\tau_2\lambda_A \\
		\ii\Delta_A^*\gamma_5\tau_2\lambda_A& 
		\ii\slashed\partial-\mu\gamma_0 
		- \gamma^\mu \omega_\mu - m_0 - \sigma
		-\ii\gamma_5{\vec \tau}^T\cdot{\vec{\pi}} 
	\end{array}\right)\,.
\end{eqnarray}
Since the functional integral over the bispinor fields is of gaussian type
the quark degrees of freedom can be integrated out and we are able to express 
the partition function in terms of collective fields only, \viz
\bea
\mathscr{Z} &=& \int
	\mathscr{D}\sigma\mathcal{D}\vec{\pi}
	\mathscr{D}\omega_\mu
	\mathscr{D}\Delta_A\mathcal{D}\Delta_A^*
	\ee^{-\int{\rm d}^4x_E~
		\left\{
		\frac{\sigma^2+\vec{\pi}^2}{4G_{\rm S}}
		-\frac{\omega_\mu^2}{4G_{\rm V}}
		+\frac{|\Delta_A|^2}{4G_{\rm D}}\right\}
		+ \frac{1}{2}\ln\det\left(\beta S^{-1}\right)}
\label{tdpot}~.
\eea
The determinant is to be taken over Dirac-, flavor-, color- and Nambu-Gorkov 
indices as well as over Euclidean 4-volume.
The latter is accomplished after Fourier transforming the inverse quark 
propagator to the momentum space and Matsubara representation given by
\bea
S^{-1} = \begin{pmatrix}
		(\ii z_n+\mu^*)\gamma_0
		-\boldsymbol{\gamma}\cdot({\bf p}
		+\boldsymbol{\omega})
		-m
		-\ii\gamma_5\vec{\tau}\cdot\vec{\pi}
		&\Delta_A \ii\gamma_5\tau_2\lambda_A
		\\
		\Delta_A^*\ii\gamma_5\tau_2\lambda_A
		&(\ii z_n-\mu^*)\gamma_0
		-\boldsymbol{\gamma}\cdot({\bf p}
		-\boldsymbol{\omega})
		-m
		-\ii\gamma_5\vec{\tau}\,^T\cdot\vec{\pi}
	\end{pmatrix}
\eea
with $z_n=(2n+1)\pi T$ being fermionic Matsubara frequencies.
Bold-face symbols denote space-like components of four-vectors and gamma 
matrices, and the transposed isospin matrices are given by 
$\vec{\tau}\,^T=(\tau_1,-\tau_2,\tau_3)$.
Finally, we have introduced the combinations
\bea
m=m_0+\sigma~,  \quad \mu^*=\mu+\omega_0~,
\eea
which can be interpreted as an effective (constituent) quark mass and an 
effective chemical potential.

\subsection{Mean-field approximation}
\label{secMeanField}	

To proceed further, we employ the homogenous mean-field approximation, \ie
we replace all occurring fields by homogeneous and isotropic mean fields.
Then the functional integration in Eq.~(\ref{tdpot}) becomes trivial, and the 
partition function essentially  factorizes into a gaussian part and a contribution 
from the inverse quark propagator. 
Accordingly, the thermodynamic potential per volume, 
can be separated into a condensate part and a contribution from the quarks,
\bea
	\Omega_{\rm MF}
	=
	\Omega_{\rm cond}
	+
	\Omega_{\rm Q}~,
\eea		
with
\bea	
	\Omega_{\rm cond}
	=
	\frac{\sigma_{\rm MF}^2}{4G_{\rm S}}
	+\frac{|\Delta_{\rm MF}|^2}{4G_{\rm D}}
	-\frac{\omega_{\rm MF}^2}{4G_{\rm V}}
\label{Omega_cond}
\eea	
and
\bea	
	\Omega_{\rm Q}
	=	
	- \frac{1}{2}\frac{T}{V}
	  \Tr\ln \left(\beta S_{\rm MF}^{-1}\right)~.
\eea
Here the functional trace symbol $\Tr$ stands for summation over  
Matsubara frequencies and three-momenta as well as
for the trace over internal  degrees of freedom, 
\ie color, flavor, Dirac and Nambu-Gorkov space.
The extra factor of $\frac{1}{2}$ arises from 
the artificial doubling of the degrees of freedom in Nambu-Gorkov formalism.

The isotropy implies $\boldsymbol{\omega}_{\rm MF}=0$, hence $\omega_{\rm MF}$ 
denotes the 0-th component only.
Also, anticipating that pseudoscalar mean fields are disfavored at nonzero bare
quark masses and vanishing isospin chemical potential, $\vec\pi_{\rm MF}$ 
vanishes as well, and therefore was dropped in Eq.~(\ref{Omega_cond}).
Moreover, we have taken the freedom to perform a global color rotation, so 
that in the diquark sector the only non-vanishing mean fields correspond to 
the $A=2$ direction, meaning that only the first two quark colors (red and 
green) participate in the condensate, while the third color (blue) remains 
unpaired.

The inverse quark propagator then takes the form
\bea
S_{\rm MF}^{-1} = \begin{pmatrix}
		(\ii z_n+\mu^*)\gamma_0
		-\boldsymbol{\gamma}\cdot{\bf p}
		-m
		&\Delta_{\rm MF}\ii\gamma_5\tau_2\lambda_2
		\\
		\Delta_{\rm MF}^*\ii\gamma_5\tau_2\lambda_2
		&(\ii z_n-\mu^*)\gamma_0
		-\boldsymbol{\gamma}\cdot{\bf p}
		-m
	\end{pmatrix}
	~.
\label{gapMF}
\eea
The inverse propagator can be inverted to obtain the propagator
\bea
S_{\rm MF} \equiv
	\left(
	\begin{smallmatrix}
		{\bf G}^+ & {\bf F}^- \\
		{\bf F}^+ & {\bf G}^-
	\end{smallmatrix}
	\right)~.
\eea
The resulting normal and anomalous Nambu-Gorkov components are
\begin{eqnarray}
{\bf G}^\pm_{\bf p} &=&
	\sum_{s_p}\Bigg(\sum_{t_p}
	\frac{t_p}{2E_{\bf p}^{\pm s_p}}
	\frac{t_pE_{\bf p}^{\pm s_p}-s_p\xi_{\bf p}^{\pm s_p}}
		 {\ii z_n-t_pE_{\bf p}^{\pm s_p}}\mathcal{P}_{rg}
	+
	\frac{1}{\ii z_n+s_p\xi_{\bf p}^{\pm s_p}}\mathcal{P}_{b}
	\Bigg)\Lambda_{\bf p}^{-s_p}\gamma_0\,,
\label{Gpm}
	\\
	{\bf F}^\pm_{\bf p}
	&=&
	\ii\sum_{s_p,t_p}\frac{t_p}{2 E_{\bf p}^{\pm s_p}}
	\frac{\Delta_{\rm MF}^\pm}{\ii z_n-t_pE_{\bf p}^{\pm s_p}}
	\tau_2\lambda_2\Lambda_{\bf p}^{s_p}\,\gamma_5\,,
\end{eqnarray}
where $s_p,t_p = \pm 1$ and
$(\Delta^+_{\rm MF},\Delta^-_{\rm MF})
	\equiv(\Delta_{\rm MF}^*,\Delta_{\rm MF})$. 
$\mathcal{P}_{rg}={\rm diag}_c(1,1,0)$, and 
$\mathcal{P}_{b}={\rm diag}_c(0,0,1)$ are color projection operators, and 
$\Lambda_{\bf p}^\pm=
	[1\pm\gamma_0(\boldsymbol{\gamma}\cdot{\bf p}+m)/E_{\bf p}]/2$ 
are projectors on positive and negative energy states, respectively.
The corresponding quasi-particle dispersion relations are 
\bea
\xi_{\bf p}^\pm=E_{\bf p}\pm\mu^*
\eea
for the blue quarks and
\bea
 E_{\bf p}^\pm=\sqrt{(\xi_{\bf p}^\pm)^2+|\Delta_{\rm MF}|^2}
\eea
for the red and green quarks,  where $E_{\bf p}=\sqrt{|{\bf p}|^2+m^2}$.

The mean-field values are obtained as stationary points of the thermodynamic 
potential, \ie
\bea
\label{gapEquationGeneral}
0 = \frac{\partial \Omega_{\rm MF}}{\partial\sigma_{\rm MF}}
	&=&
	\phantom{-}\frac{\sigma_{\rm  MF}}{2G_{\rm S}}
- \frac{1}{2} \frac{T}{V}
\Tr \left(S_{\rm MF}\frac{\partial S_{\rm MF}^{-1}}{\partial\sigma_{\rm MF}}
	\right) ~,
\\
0 = \frac{\partial \Omega_{\rm MF}}{\partial\omega_{\rm MF}}
	&=&
	-\frac{\omega_{\rm  MF}}{2G_{\rm V}}
	-\frac{1}{2}\frac{T}{V}
\Tr\left(S_{\rm MF}\frac{\partial S_{\rm MF}^{-1}}{\partial\omega_{\rm MF}}
	\right) ~,
\\	
0 = \frac{\partial \Omega_{\rm MF}}{\partial\Delta^*_{\rm MF}}
	&=&
	\phantom{-}\frac{\Delta_{\rm  MF}}{4G_{\rm D}}
-\frac{1}{2}\frac{T}{V}
\Tr\left( S_{\rm MF}\frac{\partial S_{\rm MF}^{-1}}{\partial\Delta^*_{\rm MF}}
	\right) ~,	
\\	
0 = \frac{\partial \Omega_{\rm MF}}{\partial\Delta_{\rm MF}}
	&=&
	\phantom{-}\frac{\Delta^*_{\rm  MF}}{4G_{\rm D}}
-\frac{1}{2}\frac{T}{V}
\Tr\left(S_{\rm MF}\frac{\partial S_{\rm MF}^{-1}}{\partial\Delta_{\rm MF}}
	\right) ~.	
\eea
The derivatives of the inverse propagator basically reduce to the vertex 
functions,
\bea
\frac{\partial S_{\rm MF}^{-1}}{\partial\sigma_{\rm MF}} =
	-\begin{pmatrix}
	1 & 0\\
	 0& 1
	\end{pmatrix}
	~, \quad	&&
	\frac{\partial S_{\rm MF}^{-1}}{\partial\omega_{\rm MF}}
	=
	\begin{pmatrix}
	\gamma^0 & 0\\
	 0& -\gamma^0
	\end{pmatrix}
	~, \\	
	\frac{\partial S_{\rm MF}^{-1}}{\partial\Delta^*_{\rm MF}}
	=
	\begin{pmatrix}
	 0 & 0\\
	  \ii\gamma_5\lambda_2\tau_2 & 0
	\end{pmatrix}
	~, \quad	&&
	\frac{\partial S_{\rm MF}^{-1}}{\partial\Delta_{\rm MF}}
	=
	\begin{pmatrix}
	 0 & \ii\gamma_5\lambda_2\tau_2 \\
	 0 & 0
	\end{pmatrix}
	~, \quad	
\eea
and after performing the trace in Nambu-Gorkov space the gap equations become
\bea
\label{gapsigma}
	\sigma_{\rm MF}
	&=&
	-G_{\rm S} 
	\frac{T}{V}
	\tr\left(G_{\bf p}^++G_{\bf p}^-\right)
	~,\\
	\omega_{\rm MF}
	&=&
	-G_{\rm V}
	\frac{T}{V}
	\tr\left[\left(G_{\bf p}^+-G_{\bf p}^-\right)\gamma_0\right]~,
	\\
	\Delta_{\rm MF}
	&=&
	2G_{\rm D}
	\frac{T}{V}
	\tr\left[F_{\bf p}^-
	\ii\gamma_5\tau_2\lambda_2\right]~,
	\\
\label{gapDeltas}	
	\Delta^*_{\rm MF}
	&=&
	2G_{\rm D}
	\frac{T}{V}
	\tr\left[F_{\bf p}^+
	\ii\gamma_5\tau_2\lambda_2\right]~,
\eea
where $\tr$ denotes the remaining functional trace.
Carrying out the trace in color, flavor and Dirac space and performing
the sum over  Matsubara frequencies using
\bea
T\sum_{n} \frac{1}{iz_n - x} = n(x)~,
\eea
with $n(x)=[\exp(x/T)+1]^{-1}$ being the Fermi distribution function, 
we finally obtain
\bea
	\label{gapMass}
	m-m_0 
	&=& 
	4N_fG_{\rm S}m
	\int\frac{{\rm d}^3p}{(2\pi)^3}~
	\frac{1}{E_{\bf p}}
	\left\{
		\left[1 - 2n(E_{\bf p}^-)\right]
		\frac{\xi_{\bf p}^-}{E_{\bf p}^-} 
		+\left[1 - 2n(E_{\bf p}^+) 
	\right]
	\frac{\xi_{\bf p}^+}{E_{\bf p}^+}+ n(-\xi_{\bf p}^+) - n(\xi_{\bf p}^-)
	\right\}~,
	\\
	\mu-\mu^*
	&=&
	4N_fG_{\rm V}
	\int\frac{{\rm d}^3p}{(2\pi)^3}~
	\left[
		[1-2 n(E_{\bf p}^+)]\frac{\xi_{\bf p}^+}{E_{\bf p}^+}
		-
		[1-2 n(E_{\bf p}^-)]\frac{\xi_{\bf p}^-}{E_{\bf p}^-}
		+
		n(\xi_{\bf p}^-)
		-
		n(\xi_{\bf p}^+)
	\right]~,	
\\
	\label{gapDelta}
	\Delta_{\rm MF}
	&=& 
	4N_fG_{\rm D}\Delta_{\rm MF}
	\int\frac{{\rm d}^3p}{(2\pi)^3}~
	\left[
		\frac{1-2n(E_{\bf p}^-)}{E_{\bf p}^-} 
		+ 
		\frac{1-2n(E_{\bf p}^+)}{E_{\bf p}^+}
	\right]~.
\eea
For $\Delta_{\rm MF}^*$ one gets just the complex conjugate of 
Eq.~(\ref{gapDelta}).
Moreover, only the modulus of $\Delta_{\rm MF}$ is fixed by the gap equations, 
while the choice of the phase is arbitrary. 
In practice, $\Delta_{\rm MF}$ is therefore usually chosen to be real. 
However, the complex nature of the diquark field must be kept in mind when 
fluctuations are taken into account.

The corresponding thermodynamic potential is then readily evaluated to
\bea
	\Omega_{\rm MF}
	&=&
	\frac{\sigma_{\rm MF}^2}{4G_{\rm S}}
	-
	\frac{\omega_{\rm MF}^2}{4G_{\rm V}}
	+
	\frac{|\Delta_{\rm MF}|^2}{4G_{\rm D}}
	-
	2N_f\int\frac{{\rm d}^3p}{(2\pi)^3}~
	\left[
		E_{\bf p}^+ + 2T\ln(1+e^{-E_{\bf p}^+/T})\right.\nn
		+
		E_{\bf p}^- + 2T\ln(1+\ee^{-E_{\bf p}^-/T})
		\\
		&&\left. 
		+ E_{\bf p} 
		+T\ln(1+\ee^{-\xi_{\bf p}^+/T})
		+T\ln(1+\ee^{-\xi_{\bf p}^-/T})
	\right]
	~.
\eea
In general the gap equations have more than one solution. 
The stable solution is then the set of selfconsistent  mean fields which 
minimizes the thermodynamic potential.
For the standard choice of attractive scalar quark-antiquark and quark-quark 
interactions and repulsive vector interactions, this solution corresponds to a 
minimum w.r.t.\@ $\sigma_{\rm MF}$ and $\Delta_{\rm MF}$, but to a maximum 
w.r.t.\@ $\omega_{{\rm MF}}$. 
The latter is just a constraint for thermodynamic consistency.

The above expressions get strongly simplified in the non-superconducting 
(``normal'') phase, where the diquark condensates vanish, 
$\Delta_{\rm MF} = 0$. 
In this case, the remaining gap equations reduce to
\bea
	m-m_0
	&=&
	4N_fN_cG_{\rm S}
	\int\frac{{\rm d}^3p}{(2\pi)^3}~
	\frac{m}{E_{\bf p}}
	\left[
		1-n(\xi_{\bf p}^+)-n(\xi_{\bf p}^-)	
	\right]
	~,
	\\
	\mu-\mu^*
	&=&
	4N_fN_cG_{\rm V}
	\int\frac{{\rm d}^3p}{(2\pi)^3}~
	\left[
		n(\xi_{\bf p}^-)
		-
		n(\xi_{\bf p}^+)
	\right]
	~,
\eea
while the thermodynamic potential becomes
\bea
	\Omega_{\rm MF}
	&=&
	\frac{\sigma_{\rm MF}^2}{4G_{\rm S}}
	-
	\frac{\omega_{\rm MF}^2}{4G_{\rm V}}
	-2N_fN_c
	\int\frac{{\rm d}^3p}{(2\pi)^3}~
	\left[
	E_{\bf p} + T\ln\left(1+\ee^{-(E_{\bf p}-\mu^*)/T}\right)
	+
	T\ln\left(1+\ee^{-(E_{\bf p}+\mu^*)/T}\right)
	\right]
	~.
\eea
The prefactors $N_f,N_c$ are obtained naturally from the color and flavor 
traces and are a consequence of persisting isospin and color symmetry.

\subsection{Beyond mean field: gaussian approximation}
\label{ssec:BMF}

In mean-field approximation the thermodynamic potential corresponds to a
Fermi gas of quasi-particles with dispersion relations which could be strongly
modified compared to the non-interacting case by the various mean fields.
On the other hand, the effects of low-lying bosonic excitations, in particular 
of the Goldstone bosons of the spontaneously broken symmetries,  are completely
missing. 
This could be a rather bad approximation at low temperatures, where 
the excitation of the fermionic quasi-particles is strongly suppressed by large
constituent quark masses or pairing gaps, so that the Goldstone bosons are the
dominant degrees of freedom.

In this section we therefore allow for fluctuations of the meson and diquark 
fields around their mean-field values and derive their contributions to the 
thermodynamic potential.
In this article, we will fix the order parameters on the mean-field level only,
\ie by the gap equations derived in the previous subsection.
Eventually, one should derive the generalized gap equations, where the 
fluctuation corrections to the mean-field thermodynamic potential are taken 
into account in the minimization procedure. 

We slightly change our notation in the $\sigma$ and $\omega_0$ channels and
introduce shifted fields,
\bea
\sigma &\rightarrow& \sigma_{\rm MF} + \sigma~,
\\
\omega_0 &\rightarrow& \omega_{\rm MF} + \omega_0~,               
\eea
so that, from now on, $\sigma$ and $\omega_0$ denote only the fluctuating parts
of the fields. 
Obviously, this is also true for the meson fields with vanishing mean fields, 
\ie pions and the space-like components of the $\omega$. 
For the diquarks we write

\bea
\Delta_2  = \Delta_{\rm MF} + \delta_2~,
\quad
\Delta_{5}  \equiv  \delta_{5}~,
\quad
\Delta_{7}  \equiv  \delta_{7}~,
\eea

and analogously for the complex conjugate fields. 

The gaussian terms in the partition function, Eq.~(\ref{tdpot}), are then to be
replaced by
\bea
        \sigma^2 + \vec\pi\,^2 
        &\rightarrow& 
        \sigma_{\rm MF}^2 + 2\sigma_{\rm MF}\sigma + \sigma^2 + \vec\pi\,^2~,
        \\
        \omega_\mu^2  
        &\rightarrow& 
        \omega_{\rm MF}^2 + 2 \omega_{\rm MF} \omega_0 + \omega_\mu^2 ~,
        \\
        |\Delta_A|^2  
        &=& 
        |\Delta_{\rm MF}|^2 
        + \Delta_{\rm MF}\delta_2^* +\Delta_{\rm MF}^*\delta_2 
        + |\delta_A|^2~,
\eea
where a sum over the indices $\mu$ and $A$ was implied, as before.

In addition, the fluctuations contribute to the partition function via the 
inverse propagator, which can be written as
\bea
        S^{-1} = S^{-1}_{\rm MF} + \Sigma~, 
\eea
with 
\bea
	\Sigma
	&=&
	\begin{pmatrix}
		\gamma^\mu\omega_\mu
		-\sigma
	         -\ii\gamma_5\vec{\tau}\cdot\vec{\pi}
	&\delta_A\ii\gamma_5\tau_2\lambda_A\\
	\delta_A^*\ii\gamma_5\tau_2\lambda_A
	&	-\gamma^\mu\omega_\mu
          	-\sigma
		-\ii\gamma_5\vec{\tau}\,^T\cdot\vec{\pi}
	\end{pmatrix}~.
\eea
The logarithmic term in Eq.~(\ref{tdpot}) then takes the form
\bea
	\ln\det\left(\beta S^{-1}\right)
	&=&
	\Tr\ln\left(\beta S_{\rm MF}^{-1}+\beta\Sigma\right)
	=
	\Tr\ln \left(\beta S_{\rm MF}^{-1}\right) 
	+ \Tr\ln(1+S_{\rm MF}\Sigma)
\eea
and can be Taylor-expanded in the fluctuating fields.
Here we expand it up to quadratic (gaussian) order as to account for 
two-particle correlations.
Higher-order correlations will be ignored and are left to a later 
investigation.
Those correlations would include baryons.
We obtain
\bea
	\ln\det\left(\beta S^{-1}\right) 
	=
	\Tr\ln\left(\beta S_{\rm MF}^{-1}\right)
	+
\Tr\left(S_{\rm MF}\Sigma-\frac{1}{2}S_{\rm MF}\Sigma S_{\rm MF}\Sigma\right)
	+ \mathscr{O}\left( \Sigma^3 \right)
	~,
\eea
and the partition function in gaussian approximation neglecting 
contributions $\mathscr{O}\left( \Sigma^3 \right)$ can be written as
\bea
	\mathscr{Z}_{\rm Gau{\ss}}
	&=&
	\mathscr{Z}_{\rm MF} 
	\int
	\mathscr{D}\sigma\mathscr{D}\vec{\pi}
	\mathscr{D}\omega_\mu
	\mathscr{D}\delta_A\mathscr{D}\delta_A^*
	\ee^{\mathscr{A}^{(1)}+ \mathscr{A}^{(2)}}
\eea
with the mean-field partition function 
$\mathscr{Z}_{\rm MF} = \exp \left(-\beta V \Omega_{\rm MF}\right)$,
and two correction terms to the Euclidean action which are linear 
and quadratic in the fluctuations, \viz
\bea
\mathscr{A}^{(1)}
	&=&
	-\beta V \left( \frac{\sigma_{\rm MF}\sigma}{2G_{\rm S}}
	+
	\frac{\Delta_{\rm MF}\delta_2^*+\Delta_{\rm MF}^*\delta_2}{4G_{\rm D}}
	-\frac{\omega_{\rm MF}\omega_0}{2G_{\rm V}} \right)
	+ \frac{1}{2} \Tr \left[S_{\rm MF}\Sigma\right]~,
\label{Omega_1}
\eea
and
\bea
	\mathscr{A}^{(2)}
	&=&
	-\beta V \left(\frac{\sigma^2+\vec{\pi}^2}{4G_{\rm S}}
	+\frac{|\delta_A|^2}{4G_{\rm D}}
	-\frac{\omega_\mu^2}{4G_{\rm V}}\right)
	-\frac{1}{4} \Tr\left[S_{\rm MF}\Sigma S_{\rm MF}\Sigma\right]~.
\label{Omega_2}
\eea
The linear correction term vanishes as a result of the mean-field gap 
equations. 
To show this, we evaluate the  Nambu-Gorkov trace of the last term in 
Eq.~(\ref{Omega_1}),
\bea
	-\Tr\left[S_{\rm MF}\Sigma\right]
	=
	\frac1 2\tr \left[
		G_{\bf p}^+ (\sigma +\ii\gamma_5\vec{\tau}\cdot\vec{\pi}
                                         -\gamma^\mu\omega_\mu)
		+G_{\bf p}^-(\sigma+\ii\gamma_5\vec{\tau}\,^T\cdot\vec{\pi}
                                         +\gamma^\mu\omega_\mu)\right.
\nonumber \\                                         
                   \left.
		-F_{\bf p}^+\delta_A\ii\gamma_5\tau_2\lambda_A
		-F_{\bf p}^-	\delta_A^*\ii\gamma_5\tau_2\lambda_A
	\right]~.
\eea
The pion fields and the space-like $\omega$-components drop out
in the subsequent flavor trace and momentum integral, respectively, as the
normal propagator components, given in Eq.~(\ref{Gpm}), respect the isospin symmetry and 
the isotropy of the medium.
Similarly, the $A=5,7$ diquark fields drop out in the color trace.
The remaining terms are cancelled by the  other terms in Eq.~(\ref{Omega_1}),
which can be seen easily when we group them together field-wise and use the 
gap equations~(\ref{gapsigma}) -- (\ref{gapDeltas}) to obtain 
\bea
	\frac1 2\left(\frac{\sigma_{\rm MF}}{G_{\rm S}}
	+\frac{T}{V}
	\tr\left(G_{\bf p}^++G_{\bf p}^-\right)\right)\sigma
	&=&
	0
	\\
	\frac1 2\left(
	-\frac{\omega_{\rm MF}}{G_{\rm V}}
	-\frac{T}{V}
	\tr \left[\left(G_{\bf p}^+-G_{\bf p}^-\right)\gamma_0\right]
	\right)\omega_0
	&=&
	0
	\\
	\frac1 2\left(\frac{\Delta_{\rm MF}}{2G_{\rm D}}
	-\frac{T}{V}
	\tr \left[F_{\bf p}^-\ii\gamma_5\tau_2\lambda_2
	\right]
	\right)\delta_2^*
	&=&
	0
	\\
	\frac1 2\left(\frac{\Delta_{\rm MF}^*}{2G_{\rm D}}
	-\frac{T}{V}
	\tr \left[F_{\bf p}^+\ii\gamma_5\tau_2\lambda_2
	\right]
	\right)\delta_2
	&=&
	0~.
\eea
Hence, the partition function reads
\bea
	\mathscr{Z}_{\rm Gau{\ss}}
	&=&
	\mathscr{Z}_{\rm MF}
	\int
	\mathcal{D}\sigma\mathcal{D}\vec{\pi}
	\mathcal{D}\omega_\mu
	\mathcal{D}\delta_A\mathcal{D}\delta_A^*
	\ee^{\mathscr{A}^{(2)}	
	}~.
\eea
By construction, the exponent is bilinear in the fields, so that the path 
integrals can be carried out. 
To that end, we combine all fields in a vector  
\bea
\label{X}
        X = 
        \begin{pmatrix}
         \vec\pi \\ \omega_\mu \\ \sigma \\ \delta_A \\ \delta_A^*
        \end{pmatrix}~,\quad
        X^\dagger = 
        (\vec\pi\,^T,  \omega_\mu, \sigma, \delta_A^*,  \delta_A )~,
        \label{polarizationBasis}
\eea
where all vector-meson components $\mu$ and all diquark fields, $A=2,5,7$, are 
implied, as it was in the partition function.
The trace in the exponent can then be written as 
\bea
\frac{1}{2}\Tr [S_{\rm MF}\Sigma S_{\rm MF}\Sigma]=-X^\dagger \Pi X~,
        \label{mixedPropagator}
\eea
with the polarization matrix $\Pi$.
In addition, we have the gaussian terms in front of the trace,  which are 
diagonal in this basis.
Combining both parts, we write
\bea
   	\frac{\sigma^2+\vec{\pi}^2}{2G_{\rm S}}
	+\frac{|\delta_A|^2}{2G_{\rm D}}
	-\frac{\omega_\mu^2}{2G_{\rm V}}    
    	+\frac{1}{2}\frac{T}{V}\Tr [S_{\rm MF}\Sigma S_{\rm MF}\Sigma] 
    = 
    X^\dagger {\tilde S}^{-1} X~,
\eea
with a, in general non-diagonal, propagator matrix $\tilde S$.
The partition function is then readily evaluated as  
\bea
	\mathscr{Z}_{\rm Gau{\ss}}
	&=&
	\mathscr{Z}_{\rm MF}
	\int
	\mathscr{D}X
	\ee^{-\frac1 2\int{\rm d}^4x_E~
		\left\{		
			X^\dagger {\tilde S}^{-1} X
		\right\}	
	}
	=
	\mathscr{Z}_{\rm MF}
	\left[\det \left(\beta^2 \tilde S^{-1}\right)\right]^{-1/2}~.
\eea
Note that for evaluating the gaussian functional integrals over the set of 
bosonic fields forming the components of the vector $X$ we performed a Fourier 
transformation of these fields to their representation in the space of 
three-momenta and Matsubara frequencies where they are normalized to be 
dimensionless, see chapter 2.3 of \cite{Kapusta}. 
After diagonalizing ${\tilde S}^{-1}$, the partition function factorizes,
\bea
	\mathscr{Z}_{\rm Gau{\ss}}
	=
	\mathscr{Z}_{\rm MF}
	\prod_{\rm X} \mathscr{Z}_{\rm X}~, 
\eea
with $\mathscr{Z}_{\rm X}=
	\left[\det \left(\beta^2 S_X^{-1}\right)\right]^{-d_X/2}$ 
being the partition function related to the propagator $S_{\rm X}$ of the 
$d_X-$fold degenerate eigenmode $X$ operating in the $d_X-$ dimensional 
subspace of the general propagator $\tilde S$ of two-quark correlations. 
Accordingly, the thermodynamic potential becomes a sum of the mean-field part 
and the fluctuation parts related to these modes,
\bea
	\Omega_{\rm Gau{\ss}}
	=
	-\frac{T}{V}\ln\mathscr{Z}_{\rm Gau{\ss}}
	=
	\Omega_{\rm MF}
	+
	\Omega^{(2)}
	=
	\Omega_{\rm cond}
	+
	\Omega_{\rm Q}
	+
	\sum_{\rm X} \Omega_{\rm X} ~,
\eea
with
\bea
\label{th-pot_X}
	\Omega_{\rm X}(T,\mu)
	=
	\frac{d_X}{2}\frac{T}{V}
	\Tr\ln 
	\left[\beta^2 S_{\rm X}^{-1}(\ii z_n,{\bf q})\right]
	~,
\eea
where $\Tr$ stands for summation over 3-momenta and Matsubara frequencies of
the bosonic two-particle correlation in the channel $\rm X$.

The elements of the polarization matrix $\Pi$ are explicitly listed in 
App.~\ref{App:Integrals} for the simplified case where the vector fields 
$\omega_\mu$ have been neglected.
In the 2SC phase, \ie for  $\Delta_{\rm MF} \neq 0$, the $\sigma$-meson mode 
mixes with the diquark modes $\delta_2$ and $\delta_2^*$, formally evident 
from the occurrence of non-diagonal elements in the polarization matrix.
Physically, this reflects the non-conservation of baryon number in the 
superfluid medium. 

In the non-superconducting phase where $\Delta_{\rm MF}=0$, 
on the other hand, baryon number is conserved and meson, diquark and 
anti-diquark modes decouple,
\bea
\Omega_{\rm Gau{\ss}}=
	\Omega_{\rm cond}
	+
	\Omega_{\rm Q}
	+
	\Omega_{\rm M}
	+
	\Omega_{\rm D}
	+
	\Omega_{\rm \bar D}
	~.
\eea
The three diquark modes $D=\delta_A$, related to $A=2,5,7$, 
are degenerate because of the 
persisting color symmetry, whereas the chemical potential causes a splitting 
between diquarks and anti-diquarks,
$\Omega_{\rm \bar D}(T,\mu) = \Omega_{\rm  D}(T,-\mu)$.
In the meson sector, the pions always decouple from the scalar and vector modes
because of parity and angular momentum conservation. 
On the other hand, the $\omega_0$-mode can mix with the $\sigma$ at nonzero 
$\mu$.

Neglecting again the vector fields, all elementary meson and diquark modes of 
the original Lagrangian decouple in the normal phase, 
\bea
\Omega_{\rm Gau{\ss}} =
	\Omega_{\rm cond}
	+
	\Omega_{\rm Q}
	+
	\Omega_{\sigma}
	+
	\Omega_{\pi}
	+
	\Omega_{\rm D}
	+
	\Omega_{\rm \bar D}
	~,
\label{Omegatotn}
\eea
and the correlation contributions from the composite fields are determined by
\bea
	\Omega_{\rm X}(T,\mu)
	&=&
	\frac{d_{\rm X}}{2} 
	T\sum_n \int\frac{{\rm d}^3q}{(2\pi)^3}
	\ln \left[\beta^2 S_{\rm X}^{-1}(\ii z_n,{\bf q})
	\right]\nonumber\\
	S_{\rm X}^{-1}(\ii z_n,{\bf q})
		&=&\frac{1}{G_{\rm X}}-\Pi_{\rm X}(\ii z_n, {\bf q})~.
\label{OmegaXn}	
\eea
Here $G_{\rm X}$ is the coupling, related to the channel $\rm X$, \ie 
$G_{\rm X} = 2G_{\rm S}$ for $\rm X = \sigma, \pi$ and
$G_{\rm X} = 2G_D$ for $\rm X = \delta_A, \delta_A^*$, and
$\Pi_{\rm X}$ denotes the diagonal element of the polarization matrix  
in this channel.
The corresponding degeneracy factors $d_{\rm X}$ are 
$d_\sigma = 1$ for the sigma meson and $d_{\pi}=d_D=d_{\bar{D}}=3$ for pions, 
diquarks and anti-diquarks, respectively.

\section{Meson and diquark spectra and their mixing in the 2SC phase}
\label{sec:spectra}

In this section, we want to discuss the meson and diquark spectra, which are 
given by the poles of the propagator matrix, \ie by the zeroes of  
$\det \tilde S^{-1}$.
In the normal phase, this is simplified by the fact that the various meson and 
diquark modes separate, as mentioned above. 
In the 2SC phase, on the other hand, the polarization matrix has non-diagonal 
elements which cause a mixing of the $\sigma$ mode with the $A=2$ diquarks and 
anti-diquarks.
Therefore, we will mainly concentrate on this mixing.

The relevant piece of the inverse propagator matrix has the form
\bea
	\tilde S^{-1}_{\rm mix}
	\equiv
	\begin{pmatrix}
		S^{-1}_{\sigma\sigma}&S^{-1}_{\sigma\delta_2}
		&S^{-1}_{\sigma\delta_2^*}
		\\
		S^{-1}_{\delta_2^*\sigma}&S^{-1}_{\delta_2^*\delta_2}
		&S^{-1}_{\delta_2^*\delta_2^*}
		\\
		S^{-1}_{\delta_2\sigma}
		&S^{-1}_{\delta_2\delta_2}&S^{-1}_{\delta_2\delta_2^*}
	\end{pmatrix}
	=
	\begin{pmatrix}
		\frac{1}{2G_{\rm S}} - \Pi_{\sigma\sigma}&-\Pi_{\sigma\delta_2}
		&-\Pi_{\sigma\delta_2^*}
		\\
		-\Pi_{\delta_2^*\sigma}&\frac{1}{4G_{\rm D}} 
		-\Pi_{\delta_2^*\delta_2}
		&-\Pi_{\delta_2^*\delta_2^*}
		\\
		-\Pi_{\delta_2\sigma}
		&-\Pi_{\delta_2\delta_2}&\frac{1}{4G_{\rm D}} 
		-\Pi_{\delta_2\delta_2^*}
	\end{pmatrix}~,
\eea
with the polarization-matrix elements given in App.~\ref{App:Integrals}.
These matrix elements and therefore also the elements of the 
inverse propagator matrix depend on an external three-momentum $\bf q$ and
an external bosonic Matsubara frequency $z_n$. 
In the following both functions will be analytically continued to the complex 
plane, replacing $i z_n$ by the complex variable $z$.

In order to determine the eigenmodes, $\tilde{S}^{-1}_{\rm mix}$ needs to be 
diagonalized.
Thereby we will restrict ourselves to mesons and diquarks which are 
at rest in the medium,  ${\bf q}=0$.
Then, as shown  in App.~\ref{App:IntegralsRestFrame}, the polarization matrix 
elements simplify dramatically.
Combining them with the diagonal coupling terms, we have
\bea
	\label{polss}
	S_{\sigma\sigma}^{-1}(z)
	&=&
	\frac{1}{2G_{\rm S}}+8I_\sigma(z)+16m^2|\Delta_{\rm MF}|^2I_4(z)
	\\
	\label{pol22}
	S_{\delta_2\delta_2}^{-1}(z)
	&=&
	-4\Delta_{\rm MF}^2I_0(z)
	\\
	\label{pol2s2}
	S_{\delta_2^*\delta_2}^{-1}(z)
	&=&
	\frac{1}{4G_{\rm D}}-2I_\Delta
	-4zI_1(z) - \left(4|\Delta_{\rm MF}|^2-2z^2\right)I_0(z)
	\\
	S_{\sigma\delta_2}^{-1}(z)
	&=&
	4m\Delta_{\rm MF}\left(zI_2(z)+2I_3(z)\right)
	~,
	\label{sdmix}
\eea
while the remaining elements of $\tilde S^{-1}_{\rm mix}$ follow from the 
symmetry relations Eqs.~(\ref{App:sym1})-(\ref{App:sym5}.)
Here we have dropped the three-momentum argument, ${\bf q}=0$, for brevity.
The constant $I_\Delta$ and the functions $I_i(z)$ are the finite-temperature
extensions of the integrals introduced in  Ref.~\cite{Ebert:2004dr} and are 
explicitly given in App.~\ref{App:IntegralsRestFrame}.

The above expressions are general and valid in all phases. 
In particular, we see that the mixing terms vanish in the normal phase, where 
$\Delta_{\rm MF} = 0$.
In the 2SC phase, where the mixing persists, it is more tedious, but 
straightforward to calculate the determinant of  $\tilde S^{-1}_{\rm mix}$. 
The corresponding eigenmodes,
\ie the zeroes of the determinant must in general be determined numerically.

An exception are the Goldstone modes in the 2SC phase, which can be found 
analytically. 
These modes are related to the spontaneous breaking of the color $SU(3)$ 
symmetry down to $SU(2)$. 
In this context we remind that the color symmetry is a global symmetry in the 
present model. In QCD, where it is a local gauge symmetry, the would-be 
Goldstone modes, which are related to the five generators of the broken 
symmetry are ``eaten'' by five gluons, giving them a non-zero Meissner mass. 
Hence, naively, one would expect that in the present model there are five 
Goldstone bosons in the 2SC phase. 
However, as shown in Ref.~\cite{Blaschke:2004cs} there are in fact only three 
Goldstone bosons, with two of them having a quadratic dispersion relation, in 
agreement with the Nielsen-Chadha theorem~\cite{Nielsen:1975hm}.
This abnormal number of Goldstone bosons can be related to the nonzero color 
charge~\cite{Blaschke:2004cs,Brauner:2007uw,Watanabe:2011ec},
which arises in the 2SC phase as a consequence of the fact that only red and 
green quarks are paired. 
(In QCD, the 2SC phase is always color neutralized by the background gluon 
field~\cite{Dietrich:2003nu}.)

In order to identify the Goldstone modes, we evaluate the matrix elements at  
$z=0$.
In this limit (and additionally choosing $\Delta_{\rm MF}$ to be a real 
quantity) we can make extensive use of the symmetry relations quoted in 
Eqs.~(\ref{App:sym1})- (\ref{App:sym5}), to show that the determinant can be 
written as
\bea
	\det \tilde S^{-1}_{\rm mix}
	&=&
	\left(S_{\delta_2\delta_2^*}^{-1} - S_{\delta_2\delta_2}^{-1}\right)
	\left[
	S_{\sigma\sigma}^{-1}
	\left(S_{\delta_2\delta_2^*}^{-1} + S_{\delta_2\delta_2}^{-1}\right)
		-
		2S_{\sigma\delta_2}^{-2}
	\right]~.
	\label{mixingDeterminant}
\eea
Evaluating Eqs.~(\ref{pol22}) and (\ref{pol2s2}) at $z=0$
and using the 2SC gap equation,
\bea
1=8G_{\rm D}I_\Delta~, 
\eea
one then finds that the first term,
$S_{\delta_2\delta_2^*}^{-1} - S_{\delta_2\delta_2}^{-1}$ 
vanishes, thus proving the existence of a Goldstone mode.

The mixing-problem gets strongly simplified if the quarks in the 2SC phase are 
strictly massless, which is the case in the chiral limit for a sufficiently 
weak diquark coupling $G_{\rm D}$. 
In this case the mixing between the $\sigma$ and the diquark and anti-diquark 
vanishes, see Eq.~(\ref{sdmix}),
\ie the mixing is restricted to the $A=2$ diquark and anti-diquark sector.  
The determinant of the corresponding mixing matrix is then given by 
\bea
	\det \tilde S^{-1}_{\rm mix, D \bar D}(z) =
	4z^2[(z^2-4\Delta_{\rm MF}^2)I_0(z)^2-4I_1(z)^2]
	~,
\label{Smixcl}
\eea
which makes the Goldstone mode at $z=0$ explicit.
In addition, it has a second root, which can be found by solving the 
selfconsistent equation
\bea
	z
	=
	2\Delta_{\rm MF}
	\sqrt{1+\bigg(\frac{I_1(z)}{\Delta_{\rm MF} I_0(z)}\bigg)^2}
	~.
\eea
This solution is manifestly above the threshold for pair breaking 
$2\Delta_{\rm MF}$ and thus unstable.

The remaining Goldstone modes are found in the other color directions of the 
diquark sector, which do not mix.
In these channels the inverse propagators read
\bea
	S_{\delta_5^*\delta_5}^{-1}(z)
	&=&
	S_{\delta_7^*\delta_7}^{-1}(z)
	=
	\frac{1}{4G_{\rm D}}-I_\Delta - 2zI_7(z)
	+(|\Delta_{\rm MF}|^2-z^2)I_5(z)
	\label{pol57}
\eea
and $S_{\delta_5\delta_5^*}^{-1}(z) =S_{\delta_7\delta_7^*}^{-1}(z)=
	S_{\delta_5^*\delta_5}^{-1}(-z)$.
Observing that $I_5(z=0)=-I_\Delta/|\Delta_{\rm MF}|^2$ and using the
gap equation again, one finds that these inverse propagators also vanish 
at $z=0$.

A closer inspection shows that $\det \tilde S^{-1}_{\rm mix}$ yields a factor 
$z^2$ (as evident in the chiral limit from Eq.~ (\ref{Smixcl})), whereas
$S_{\delta_A\delta_A^*}^{-1}(z)$ and $S_{\delta_A^*\delta_A}^{-1}$ in the
$A=5$ and $7$ sectors only contribute a factor $z$ each, so that diquarks and
anti-diquarks must be combined to be counted as a full Goldstone mode. 
This leads to a total number of three Goldstone bosons, as already mentioned 
above.

For completeness,  the pion propagator at rest is given by 
\bea
	S_{\pi\pi}^{-1}(z)
	&=&
	\frac{1}{2G_{\rm S}}+8I_\pi(z)~,
\eea
with the function $I_\pi(z)$ as defined in Eq.~(\ref{Ipi}).
As well known, in the chiral limit, the pions are the Goldstone bosons in the
chirally broken normal phase ($m\neq 0$), while for $m=0$, they become 
degenerate with the $\sigma$ meson.

Finally, we would like to point out that, in general, the pole energies of the 
eigenmodes at ${\bf q}=0$ should not be called ``masses'', although this is 
quite common in the literature \cite{Zablocki:2012zz,Strodthoff:2011tz}. 
To see this, we recall that the propagator of a free boson with mass 
$m_{\rm X}$ at boson-chemical potential $\mu_{\rm X}$ has the form
\bea
S_{\rm X}^{\rm free} = \frac{1}{(z+\mu_{\rm X})^2 - {\bf q}^2 - m_{\rm X}^2}
	~,
\label{Sbfree}
\eea
which, for ${\bf q}=0$, has poles at 
$z = \pm m_{\rm X} - \mu_{\rm X} \equiv \omega_{\rm X}^\pm$.
Thus, we should identify the mass and the chemical potential of the bosonic 
mode $\rm X$ from its pole energies as~ \cite{Kleinhaus:2007ve}
\bea
	m_{\rm X}  &=& \frac{1}{2} ( \omega_{\rm X}^+ - \omega_{\rm X}^-)
	~,
	\quad
	\mu_{\rm X} = -\frac{1}{2} ( \omega_{\rm X}^+ + \omega_{\rm X}^-)
	~.
\label{mxmux}
\eea
Of course, in the normal phase, the assignment of the chemical potentials 
corresponds to the net quark-number content of the boson, \ie 
$\mu_{\rm X} = 0$ for the mesons, 
$\mu_{\rm X} = 2\mu^*$ for the diquarks, and 
$\mu_{\rm X} = -2\mu^*$ for the anti-diquarks. 
As a consequence, the poles of the diquark and anti-diquark propagators split, 
even at $T=0$ and low chemical potentials (see, \eg Ref.~\cite{Ebert:2004dr}), 
while their true masses stay  at their vacuum values until the lowest 
excitation threshold is reached or a phase transition takes place.

In the 2SC phase, on the other hand, where baryon number is not conserved and 
mixing takes place, the chemical potentials $\mu_{\rm X}$ related to the 
various modes are less clear a priori and must be determined from 
Eq.~(\ref{mxmux}). 

\section{Generalized Beth-Uhlenbeck equation of state}
\label{sec:GBU}

In the present section we will formulate the thermodynamics of two-particle 
correlations in a form which is known as the Beth-Uhlenbeck EoS 
\cite{Beth:1936zz,Beth:1937zz}. 
The standard Beth-Uhlenbeck formula considers the second virial coefficient 
for the EoS that contains the contribution of bound states and scattering 
states in the low-density limit. 
In dense matter, the single-particle properties as well as the two-particle 
properties are modified. 
This is seen in the corresponding spectral functions where the $\delta$-like 
peaks describing the single-particle states and two-particle bound states (in 
the two-particle propagator) are shifted and broadened. 
The medium modifications of these quasiparticles are given in 
lowest approximation by the self-energy and (in the two-particle case) 
screening and Pauli-blocking contributions. 
When speaking of high densities we have in mind the occupation of phase space
measured by scalar densities which can be large even at vanishing baryon 
density (zero chemical potential). 
We will generalize the standard Beth-Uhlenbeck  approach to the situation in 
a hot and dense medium when the gap between the discrete spectrum of 
two-particle bound states and 
the scattering continuum (defining the binding energy) diminishes and finally 
vanishes so that the bound state merges the continuum. 
This generalized Beth-Uhlenbeck (GBU) EoS is applicable in a wide range of 
densities, improving the mean-field approach by including medium-modified 
two-particle correlations. 

The dissolution of composite particles into their constituents because of
the screening of interaction in a dense medium is known 
as the Mott effect in solid state physics and has also found numerous 
applications in semiconductor physics (transition from the exciton gas
 to the electron-hole liquid \cite{Zimmermann:1985ji}), plasma physics 
(pressure ionization \cite{KKER}) and nuclear physics (cluster dissociation 
due to Pauli blocking \cite{RMS}).
Here we want to apply the concept to particle physics and formulate the 
problem of hadron dissociation as a Mott effect.
As it is known that this effect should not lead to discontinuities in the 
thermodynamic functions like the density, one has to take care of the 
normalization of the spectral function of the two-particle correlations.
Whenever a bound state gets dissociated, it should leave a trace in the 
behavior of the scattering phase shift at the threshold of the continuum.
This constraint is known as Levinson's theorem II 
\cite{Levinson,Dashen:1969ep}. 
As we shall see, it can play an important role for the formulation of a 
thermodynamics of hadronic matter under extreme conditions where one of the 
key puzzles is the mechanism of hadron dissociation at the transition to
the quark-gluon plasma (QGP) or, equivalently, the problem of hadronization of 
the QGP in the course of which correlations (pre-hadrons) form in 
the vicinity of the quark-to-hadron matter transition. 
Those correlations shall play a decisive role, \eg for understanding 
the chemical freeze-out. 
In the present section we will formulate the thermodynamics in the 
Beth-Uhlenbeck form \cite{Beth:1936zz,Beth:1937zz} which allows the discussion 
of these issues in terms of two-particle correlations (2nd virial coefficient) 
as expressed by scattering phase shifts. 

While in a low-density system, the phase shifts can be regarded as measurable 
quantities which then may be used to express deviations from an ideal 
gas behaviour due to two-particle correlations in a dilute medium, the 
situation changes in a dense system.
Under extreme conditions, in particular in the vicinity of the Mott transition,
the modification of the two-particle system by the influence of the medium has
to be taken into account. 
This has been done systematically within a thermodynamic Green function 
approach \cite{Schmidt:1990} but its extension for relativistic systems 
within a field theoretic formulation with contributions 
from antiparticles, relativistic kinematics and the role of the zero-point 
fluctuations has been missing. 
Previous work in this direction has been done by H\"ufner et al. 
\cite{Hufner:1994ma} and Abuki \cite{Abuki:2006dv}, 
who have chosen to introduce the scattering phase shifts 
as arguments of the Jost representation of the complex $\cal{S}$ matrix.
Most recently, Wergieluk et al. \cite{Wergieluk:2012gd}, Yamazaki and Matsui 
\cite{Yamazaki:2012ux}, and Dubinin et al. \cite{Dubinin:2013yga} 
used a different approach where the phase shifts are  encoding 
correlations of the relativistic two-particle propagators introduced above,
which are the analogue of the $\cal{T}$ matrix of two-body scattering theory. 
Following this formalism, we thus develop here the field theoretic analogue 
of the approach by \cite{Zimmermann:1985ji, Schmidt:1990}. 

To be as transparent as possible in this basic and new contribution, 
we choose here to present derivations in the normal phase where 
$\Delta_{\rm MF}=0$, neglecting the possibility of diquark condensation.
The derivation of the more general case works in the same manner but involves
the mathematical apparatus to deal with the mixing of correlation channels
(see, e.g., \cite{Schmidt:1990}) which shall be given elsewhere.
Note that the dissolution of composite particles into their constituents
gives no discontinuities in the thermodynamic properties as long as homogeneous
systems are considered. 
However, phase instabilities are provoked if the stability criteria according 
to the second law are violated due to the contribution of two-particle 
correlations to the EoS.

\subsection{Bound states vs. continuum. Mott effect}
\label{ssec:prop-BU}

Starting point for the derivation of the 
GBU EoS is the thermodynamic potential (\ref{Omegatotn}) which separates into 
the contributions from the eigenmodes of the two-particle propagation 
(\ref{OmegaXn}) as encoded
in the two-particle propagators $S_X(\ii z_n,{\bf q})$
with the inverse 
$
S_{\rm X}^{-1}(\ii z_n,{\bf q})= G_{\rm X}^{-1}-\Pi_{\rm X}(\ii z_n,{\bf q})
$
for a generic particle/correlation ${\rm X}\in\{{\rm M},{\rm D}\}$, defined at 
the Matsubara frequencies $\ii z_n$. 
The complex function $S^{-1}_{\rm X}(z,{\bf q})$ is its 
analytic continuation into the complex $z$-plane.
$\Pi_{\rm X}(\ii z_n,{\bf q})$ is the polarisation loop in the 
corresponding channel with an analytic continuation to the complex $z$- plane, 
analogous to that of the propagator.
The complex propagator functions can be given the polar representation
\bea
	S_{\rm X}=|S_{\rm X}|\ee^{\ii\delta_{\rm X}}=S_R+\ii S_I~,
	\label{phase}
\eea
where the scattering phase shift $\delta_{\rm X}$ has been introduced as
\begin{equation}
\label{phaseshift}
\delta_{\rm X}(\omega, {\bf q})=
-\IM\ln\left[\beta^2S_{\rm X}^{-1}(\omega-\mu_X+\ii\eta,{\bf q})\right]~.
\end{equation} 
Note that in this definition we have shifted the energy argument in the
inverse propagator by $-\mu_X$ in order to exploit the symmetry properties of 
the propagator, cf. Eq.~(\ref{Sbfree}).  
For the same  reason, in the following discussion,
the (inverse) propagator and the polarization function $\Pi_{\rm X}$
should always be understood as to be evaluated at the shifted energy
$z=\omega -\mu_X +i\eta$, unless stated otherwise.
Of course, this only concerns the diquarks and anti-diquarks, as for mesons 
we have
$\mu_X = 0$ anyway.
%

The contribution of the mode $\rm X$ to the thermodynamic potential as given
in Eq.~(\ref{th-pot_X}) can be transformed using the spectral 
respresentation of the function $\ln S_{\rm X}^{-1}(\ii z_n,{\bf q})$ and the 
definition of the phase shift (\ref{phaseshift}) to obtain the form
\bea
\label{Om_bar1}
	\Omega_{\rm X}(T,\mu) 
	&=&
	\frac{d_{\rm X}}{2} T\sum_n \int\frac{{\rm d}^3q}{(2\pi)^3}
	\ln\left[\beta^2 S_{\rm X}^{-1}(\ii z_n,{\bf q})\right]~,
\nonumber\\
 	&=&-\frac{d_{\rm X}}{2} T\sum_n \int\frac{{\rm d}^3q}{(2\pi)^3}
        \int_{-\infty}^{\infty}\frac{{\rm d}\omega}{2\pi}~
	\frac{2}{\ii z_n-\omega}
	\IM\ln\left[\beta^2 S_{\rm X}^{-1}(\omega+\ii\eta,{\bf q})\right]
	~,	
\nonumber\\
	&=&
	\frac{d_{\rm X}}{2} T\sum_n \int\frac{{\rm d}^3q}{(2\pi)^3}
        \int_{-\infty}^{\infty}\frac{{\rm d}\omega}{2\pi}~
	\frac{2}{\ii z_n-(\omega-\mu_X)} \delta_X(\omega,{\bf q})~.
\eea
Performing the bosonic Matsubara summation we arrive at 
\bea
\label{Om_bar}
	\Omega_{\rm X}(T,\mu) 
 	&=&
	-d_{\rm X}\int\frac{{\rm d}^3q}{(2\pi)^3}~
	\int_{-\infty}^{\infty}\frac{{\rm d}\omega}{2\pi}~
	n_X^-(\omega)\delta_X(\omega,{\bf q})
	~,
\eea
where we have employed the Bose function with chemical potential according
to the notation
$n_X^\pm(\omega) = 1/\{\exp[(\omega\pm\mu_X)/T]-1\}$.
After splitting the $\omega$- integration into negative and nonnegative domains
and using the fact that for the  Bose distribution function 
holds 
$n_X^-(-\omega)=-[1+n_X^+(\omega)]$ 
while the phase shift (\ref{phaseshift}) is an odd 
function under the reflection $\omega \to -\omega$ as a consequence of 
$J_{X,{\rm pair}}^\pm(-\omega)= - J_{X,{\rm pair}}^\mp(\omega)$
and 
$J_{X,{\rm Landau}}^\pm(-\omega)= - J_{X,{\rm Landau}}^\mp(\omega)$,
see App.~\ref{App:PolarizationNormal} and App.~\ref{App:PolarizationNormal-D},
we arrive at
\bea
\label{GBU-n}
	\Omega_{\rm X}(T,\mu) 
 	&=&
	- d_{\rm X}\int\frac{{\rm d}^3q}{(2\pi)^3}~
	\int_{0}^{\infty}\frac{{\rm d}\omega}{2 \pi}~
	[1+n_X^-(\omega)+n_X^+(\omega)]\delta_{\rm X}(\omega, {\bf q})
	~.
\eea
Here the shift in the definition of $\delta_X$, Eq.~(\ref{phaseshift}),
was essential to have this symmetry property for diquarks and anti-diquarks 
as well.
Performing a partial integration over the energy variable in (\ref{Om_bar}) 
leads to
\bea
\label{GBU}
	\Omega_{\rm X}(T,\mu)
	=
	d_{\rm X}\int\frac{{\rm d}^3q}{(2\pi)^3}~
	\int_0^\infty\frac{{\rm d}\omega}{2 \pi}~
	\left\{
	\omega
	+T\ln\left(1-\ee^{-(\omega-\mu_{\rm X})/T}\right)
	+T\ln\left(1-\ee^{-(\omega+\mu_{\rm X})/T}\right)
	\right\}
	\frac{d\delta_{\rm X}(\omega, {\bf q})}{d\omega}
	~.
\eea
This expression still contains the divergent vacuum energy contribution. 
We remove this term in analogy to the ``no sea'' approximation which is 
customary in relativistic mean field approaches to thermodynamics, like 
the Walecka model, and arrive at  the  Beth-Uhlenbeck formula for the 
pressure $p_X(T,\mu)=-\Omega_X(T,\mu)$,
\bea
\label{GBU-p}
	p_X(T,\mu) 
	= - d_{\rm X}T\int\frac{{\rm d}^3q}{(2\pi)^3}~
	\int_0^\infty\frac{{\rm d}\omega}{2 \pi}~
	\left\{
	\ln\left(1-\ee^{-(\omega-\mu_{\rm X})/T}\right)
	+\ln\left(1-\ee^{-(\omega+\mu_{\rm X})/T}\right)
	\right\}
	\frac{d\delta_{\rm X}(\omega, {\bf q})}{d\omega}
	~.
\eea
With Eq.~(\ref{GBU-p}) 
the medium-dependent derivative of the phase shift 
$\delta_{\rm X}^\prime(\omega, {\bf q})$ has been introduced as 
a spectral weight factor for the contribution of a two-particle state $X$ 
with degeneracy factor $d_X$, depending on the three-momentum  ${\bf q}$  
and the two-particle energy $\omega$.
Eq.~(\ref{GBU-p})  
differs from the standard Beth-Uhlenbeck equation in nonrelativistic
\cite{Beth:1937zz,Beth:1936zz}  or relativistic \cite{Dashen:1969ep}
systems by the fact that the two-particle propagator and therefore also the 
phase is obtained by taking into account in-medium effects as encoded in the 
solutions of the gap equations which define the quark quasiparticle 
(meanfield) propagators entering its definition.

Note that the phase angle used here should not be confused with
an observable scattering phase shift which should be on the energy 
shell. 
Here the function $\delta_{\rm X}(\omega, {\bf q})$ is merely a convenient
parametrization of the spectral properties of the logarithm of the 
two-particle propagator $S_{\rm X}(\omega-\mu_X +i\eta, {\bf q})$.
The latter is defined by the polarization function $\Pi_X(z,{\bf q})$,
being a one-loop integral involving mean-field quark propagators and thus not 
selfconsistently determined. 
We shall come back to the issue of selfconsistency in 
a separate work. 
 
\subsection{Phase shifts and Levinson theorem for meson dissociation}

Now we want to show that the phase (\ref{phase}) can be decomposed in two
parts, corresponding to a structureless scattering continuum $\delta_c$ 
and a resonant (collective) contribution $\delta_R$ which under appropriate
conditions represents a bound state contribution. 

The following derivation holds whenever the polarization loop integral can be 
expressed in the form 
\begin{equation}
\Pi_{\rm X}(z,{\bf q})= \Pi_{X,0} + \Pi_{X,2}(z,{\bf q})~,
\end{equation}
where $\Pi_{X,0}$ is a 4-momentum independent, real number, which can 
be a function of external thermodynamic variables.
We will show that for this very general form the decomposition 
$\delta_{\rm X}=\delta_{X,c}+\delta_{X,R}$ holds,
whereby $\delta_{X,R}$ corresponds to a resonant mode which goes over to the 
real bound state at the Mott transition where the bound state energy 
meets the threshold of the continuum of scattering states. 
The latter are described by $\delta_{X,c}$, a structureless continuum  
background phase shift with a threshold to be directly identified from 
inspection of 
$\IM \Pi_{\rm X}$ as given in 
App.~\ref{App:PolarizationNormal} and \ref{App:PolarizationNormal-D}.
Here we generalize a result which has been derived for the pion and sigma 
channels before in \cite{Zhuang:1994dw} for the NJL model and in 
\cite{Wergieluk:2012gd} for the Polyakov-loop extended NJL model.

The two-particle propagator $S_{\rm X}$ in Eq.~(\ref{phaseshift}) can be given 
the form
\bea
\label{prop}
	S_{\rm X}(z-\mu_X+\ii\eta,{\bf q})
	=
	\frac{1}{G_{\rm X}^{-1}-\Pi_{X,0}-\Pi_{X,2}(z-\mu_X+\ii\eta,{\bf q})}
	=
\frac{1}{\Pi_{X,2}(z-\mu_X+\ii\eta,{\bf q})}\frac{1}{R_X(z^2,{\bf q})-1}~,
\eea
where the auxiliary function
\bea
	R_X(z^2,{\bf q})=\frac{1-G_{\rm X}\Pi_{X,0}}
			{G_{\rm X}\Pi_{X,2}(z-\mu_X+\ii\eta,{\bf q})}
\eea
has been introduced. Now obviously holds 
\bea
	\ln S_{\rm X}(z-\mu_X+\ii\eta,{\bf q})^{-1}
	=
	\ln \Pi_{X,2}(z-\mu_X+\ii\eta,{\bf q}) +\ln[R_X(z^2,{\bf q})-1]
	~,
\eea
so that with (\ref{phase})
\bea
\label{delta}
	\delta_{\rm X}(\omega,{\bf q})
	=\delta_{X,c}(\omega,{\bf q}) + \delta_{X,R}(\omega,{\bf q})
	~,
\eea
where 
\bea
\label{delta_c}
	\delta_{X,c}(\omega,{\bf q})
	&=&
	- \arctan\left(\frac{\IM \Pi_{X,2}(\omega- \mu_X+\ii\eta,{\bf q})}
		{\RE \Pi_{X,2}(\omega-\mu_X+\ii\eta,{\bf q})} \right)
	~,
\\
	\delta_{X,R}(\omega,{\bf q})
	&=&
	\arctan\left(\frac{\IM R_X(\omega^2,{\bf q})}
		{1- \RE R_X(\omega^2,{\bf q})} \right)~. 
\label{delta_R}
\eea
From this decomposition of the phase shift it becomes immediately obvious that 
$\delta_{X,R}(\omega,{\bf q})$ corresponds to the phase shift of a resonance 
at $\omega=\omega_X=\sqrt{{\bf q}^2+m_{\rm X}^2}$. 
The position of the resonance is found from the condition 
$\RE R_X(\omega^2_X)=1$, where for brevity we drop here and in the 
following derivation the argument ${\bf q}$.
At this energy holds that $\delta_{X,R}(\omega\to \omega_X)\to \pi/2$ 
since $\tan\delta_{X,R}(\omega\to \omega_X)\to \infty$.
 
Performing a Taylor expansion at the resonance position for real $\omega$
one obtains 
\bea
1- \RE R_X(\omega^2)&\approx & \underbrace{1- \RE R_X(\omega^2_X)}_{=0}
-(\omega^2-\omega_X^2)\RE \frac{d R_X(z^2)}{d z^2}\bigg|_{z=\omega_X}~,
\\
\IM R_X(\omega^2)&\approx & \IM R_X(\omega_X^2) ~.
\eea
From this follows
\bea
\frac{1- \RE R_X(\omega^2)}{\IM R_X(\omega^2)}&\approx& - (\omega^2-\omega_X^2)
\frac{\RE \frac{d R_X(z^2)}{d z^2}\big|_{z=\omega_X}}{\IM R_X(\omega_X^2)}~.
\eea
If we now define that 
\bea
\omega_X\Gamma_X = 
- \frac{\IM R_X(\omega_X^2)}{\RE \frac{d R_X(z^2)}{d z^2}\big|_{z=\omega_X}}~,
\eea
the resonant phase shift becomes
\bea
	\delta_{X,R}(\omega,{\bf q})
	=
	\arctan\left(\frac{\omega_{X}\Gamma_{X}}
			{\omega^2-\omega_X^2} \right)~,
\eea
which corresponds to the Breit-Wigner form for the spectral density in the
Beth-Uhlenbeck EoS
\bea
\label{B-W}
\frac{ d \delta_{X,R}(\omega)}{d\omega}=
\frac{2\omega \omega_{\rm X}\Gamma_{\rm X}}
{(\omega^2-\omega_X^2)^2+\omega_{\rm X}^2\Gamma_{\rm X}^2}~.
\eea
This form goes over to the spectral density of a bound state when the width
parameter $\Gamma_{\rm X}\to 0$,
\begin{equation}
\lim_{\Gamma_{\rm X}\to 0} \frac{ d \delta_{X,R}(\omega)}{d\omega} = 
\pi \left[\delta(\omega-\omega_X)+\delta(\omega+\omega_X)
\right]~,
\end{equation}
with the Dirac $\delta$ distribution on the r.h.s.

The continuum contribution is defined along a cut on the real axis in the 
complex energy plane, i.e. for 
$\omega \ge \omega_{\rm thr}({\bf q})=\sqrt{{\bf q}^2 + 4 m^2}$,
where $\IM \Pi_2\neq 0$.
The value of the corresponding phase shift at threshold vanishes,
$\delta_{X,c}(\omega_{\rm thr})=0$.

\begin{figure}[htbp] 
\vspace{5mm}
\centering{
\includegraphics[angle=0,width=0.45\textwidth]
	{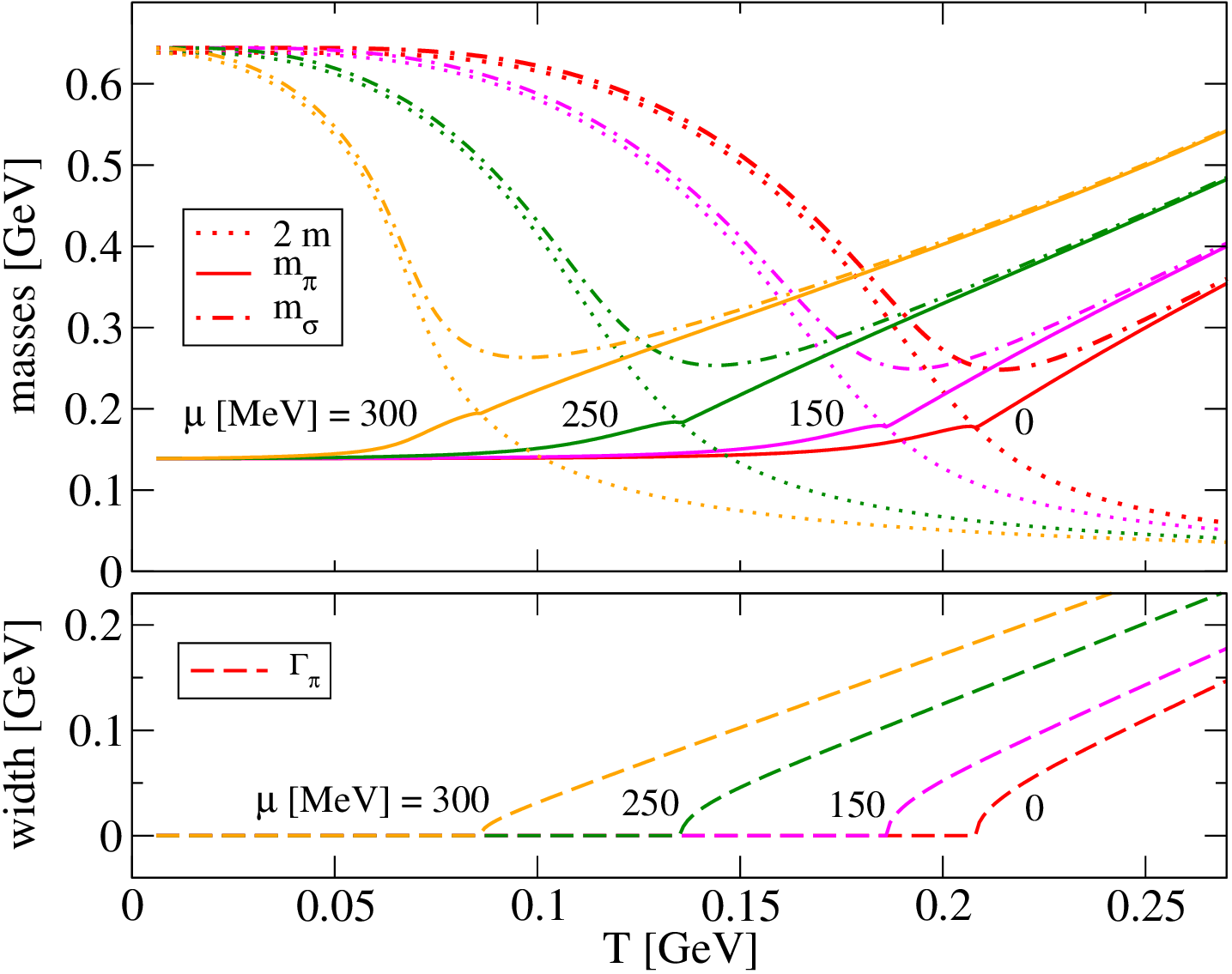}
\hspace{1cm}
\includegraphics[angle=0,width=0.45\textwidth]
	{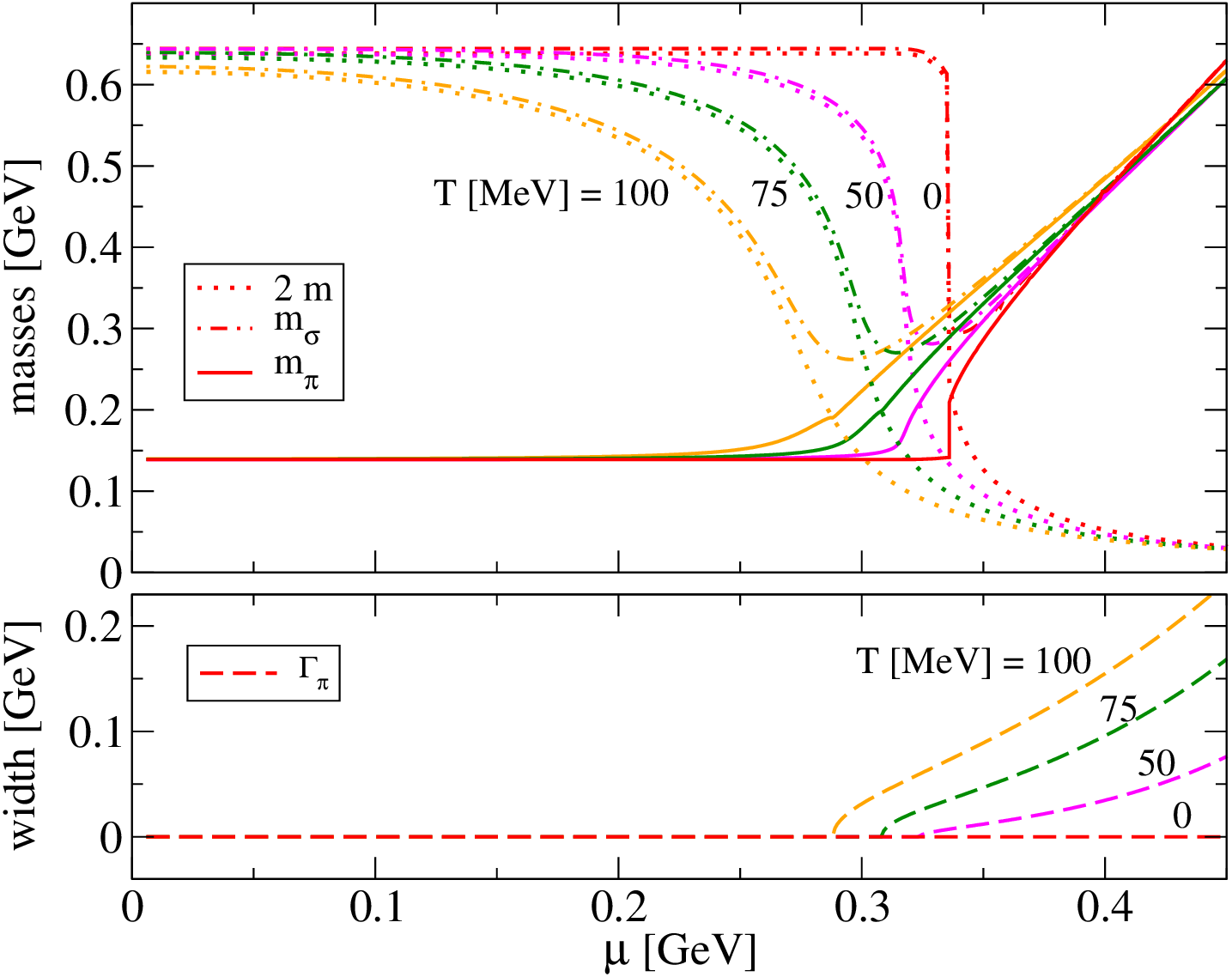}} 
\caption{(Color online).
Two-particle spectrum in the scalar-pseudoscalar meson channels at rest in the
medium (${\bf q}=0$). 
Left panel: Behaviour of the thresholds for the quark-antiquark continuum $2m$ 
(dotted lines), meson masses ($m_\sigma$ - dash-dotted lines, $m_\pi$ - 
solid lines) and pion width $\Gamma_\pi$ (dashed lines)
as functions of $T$ for different values of the chemical potential $\mu$. 
Right panel: 
Same as left panel but as functions of $\mu$ for different temperatures $T$. 
} 
\label{fig:massspectra}
\end{figure}

If the energy of the state $X$ is below that threshold, 
$\omega_X<\omega_{\rm thr}$,
it is a real bound state with vanishing width ($\Gamma_X=0$, infinite lifetime)
and the resonant phase shift behaves as a step function which jumps by $\pi$ 
at $\omega = \omega_X$  and has therefore this value at the threshold, 
$\delta_{X,R}(\omega_{\rm thr})=\pi$ for $T<T_{\rm X, Mott}$.
In Fig.~\ref{fig:massspectra} we show the behaviour of the threshold, the 
meson masses and the pion width in sections through the phase diagram in the
$T,\mu-$ plane, for ${\bf q}=0$.
Note that the sharp onset of the pion width  at the Mott temperatures
is an artefact of the present approximation which neglects 
$\pi-\pi$ scattering and treats quarks in the mean field 
approximation only, where they are on-shell quasiparticles with zero width. 
  
The parameters employed are a bare quark mass $m_0 = 5.5$~MeV, 
a three-momentum cutoff $\Lambda = 639$~MeV and a scalar coupling constant 
$G_{\rm S}\Lambda^2 = 2.134$. 
We consider two alternative values for the diquark coupling constant: 
a strong coupling $G_{\rm D}= G_{\rm S}$, and a moderate coupling 
$G_{\rm D}=3/4~G_{\rm S}$. The latter is motivated by the ratio between 
scalar diquark and scalar meson interaction channels arising from a Fierz 
transformation of the (massive) vector boson exchange interaction model
(see, e.g., Eq.~(A.20) in the App.~A of Ref.~\cite{Buballa:2003qv}). 
With the above parameters one finds in vacuum a constituent quark mass of 
319~MeV, a pion mass of 138~MeV and pion decay constant $f_\pi=92.4$ MeV. 
The vacuum mass of the  $\sigma$-meson is 644~MeV,
which is thus slightly unbound.
The scalar diquark is bound (pion-like) for the strong diquark coupling and 
slightly unbound (sigma-like) for the Fierz value of the coupling.
Masses and widths of mesons and diquarks
are determined from the solution of 
the corresponding Bethe-Salpeter equations, 
\ie from the (complex) poles of their two-particle propagators,
see Sects.~\ref{ssec:BMF}   and \ref{sec:spectra}.

\begin{figure}[htbp] 
\centering{
	\includegraphics[width=0.9\textwidth]{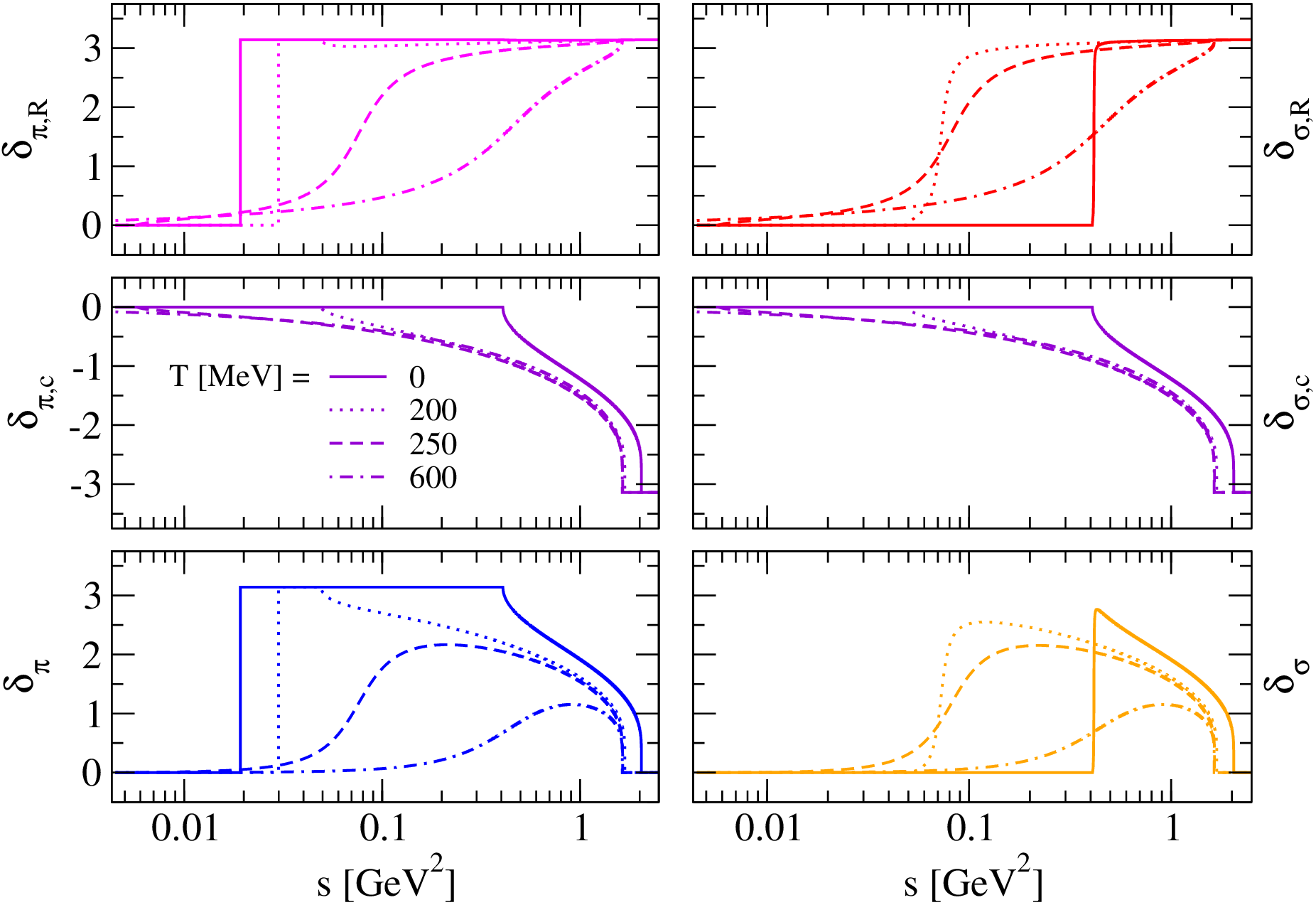}
}
\caption{
Phase shifts for pion (left panels) and sigma meson (right panels) channels 
at rest (${\bf q}=0$) as functions of the squared mass variable $s=\omega^2$ 
for different temperatures, below ($T=0,~200$ MeV) and above ($T=250,~600$ MeV)
the Mott-transition. The upper panels show the resonant phase shifts which 
together with the continuum phase shifts shown in the middle panels add up
to the total phase shifts of the bottom panels.
}
\label{Fig:Phi_matrix}
\end{figure}

The Mott transition, when the bound state merges the continuum, can be 
detected also from the behaviour of the resonant phase shift which will be 
subject to the same threshold as the continuum and vanishes at the continuum 
edge, $\delta_{X,R}(\omega_{\rm thr})=0$ for $T>T_{\rm X, Mott}$. 
This is a manifestation of {Levinson's theorem II} which can be formulated as
\bea
\label{levinson}
\int_0^{\infty}d\omega \frac{1}{\pi}\frac{d\delta_X(\omega;T)}{d\omega} = 0
= \underbrace{\int_0^{\omega_{\rm thr}(T)}d\omega \frac{1}{\pi}
\frac{d\delta_X(\omega;T)}{d\omega}}_{n_{B,X}(T)} 
+ \underbrace{\frac{1}{\pi}\int_{\omega_{\rm thr}(T)}^{\infty}
d\omega\frac{d\delta_X(\omega;T)}{d\omega}
}_{\frac{1}{\pi}[\delta_X(\infty;T)-\delta_X(\omega_{\rm thr};T)]}~,
\eea 
at any given temperature $T$.
Since under very general conditions \cite{Levinson} holds
$\delta_X(\infty;T)=0$ for any temperature it follows that
$\delta_X(\omega_{\rm thr};T)=\pi n_{B,X}(T)$,\ie that decrementing 
the number of bound states in the channel $X$ at the corresponding Mott 
temperature $T_{X, {\rm Mott}}$ has to be accompanied by a jump by $\pi$ 
of the phase shift at threshold \cite{Dashen:1969ep}.
This behaviour is illustrated in Fig.~\ref{Fig:Phi_matrix}, see also 
Refs.~\cite{Wergieluk:2012gd,Dubinin:2013yga}.

\subsection{Thermodynamics of Mott dissociation}

In this subsection, we demonstrate first for the case of the pion as the 
lightest meson and then for the scalar diquark, how the Mott dissociation of 
the bound state leads to a reduction of the correlation contribution to the 
thermodynamical potential.
All thermodynamic relations can be consistently derived from the 
thermodynamic potential which for a homogeneous system is directly given by 
the pressure, were for the following numerical 
investigation we focus on $X=\pi, \delta_A, \delta_A^*$.

\subsubsection{The case of the lightest meson}

In order to discuss the pressure for pion correlations in quark matter 
we start from the expression (\ref{GBU-p}). 
We respect that $\mu_\pi=0$ so that
the GBU EoS for the pion pressure takes the form
\bea
\label{GBU2}
p_{\pi}(T)= -d_{\pi}T\int\frac{{\rm d}^3q}{(2\pi)^3}~
	\int_0^\infty\frac{{\rm d}\omega}{\pi}~
	\ln\left(1-\ee^{-\omega/T}\right)
	\frac{d\delta_{\pi}(\omega, {\bf q})}{d\omega}	~.
\eea
This equation encodes the in-medium modification of the pionic correlations
including their Mott dissociation in the behaviour of the phase shift 
$\delta_{\pi}(\omega, {\bf q})$ which is defined by 
Eqs.~(\ref{delta})-(\ref{delta_R}) via the polarization function examined in 
detail in Appendix \ref{App:Integrals}.
For a recent numerical evaluation of the pion and sigma meson thermodynamics
within the PNJL model, see \cite{Yamazaki:2012ux}.

\begin{figure}[htbp] 
\centering{
	\includegraphics[width=0.48\textwidth]{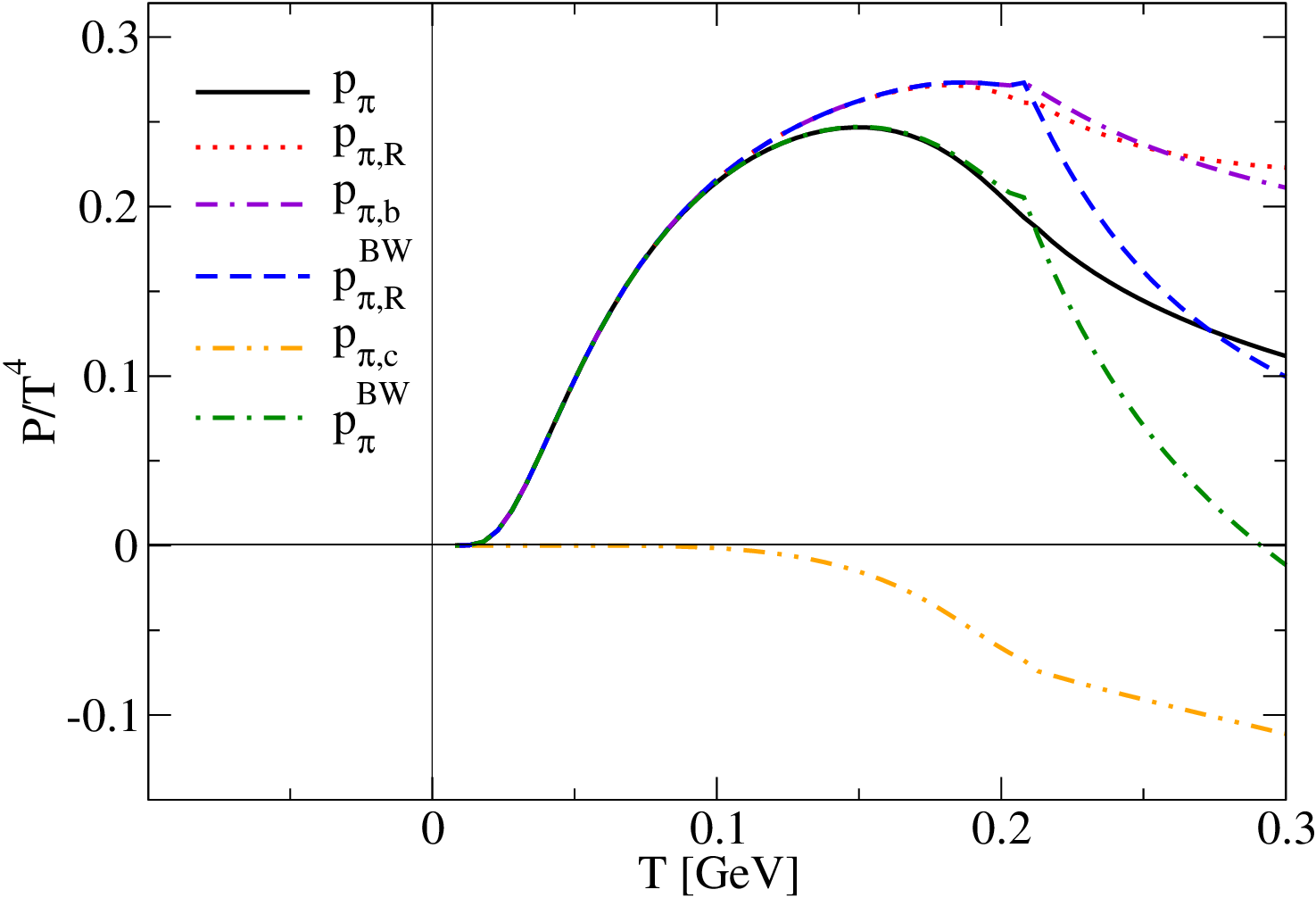}
	\includegraphics[width=0.48\textwidth]{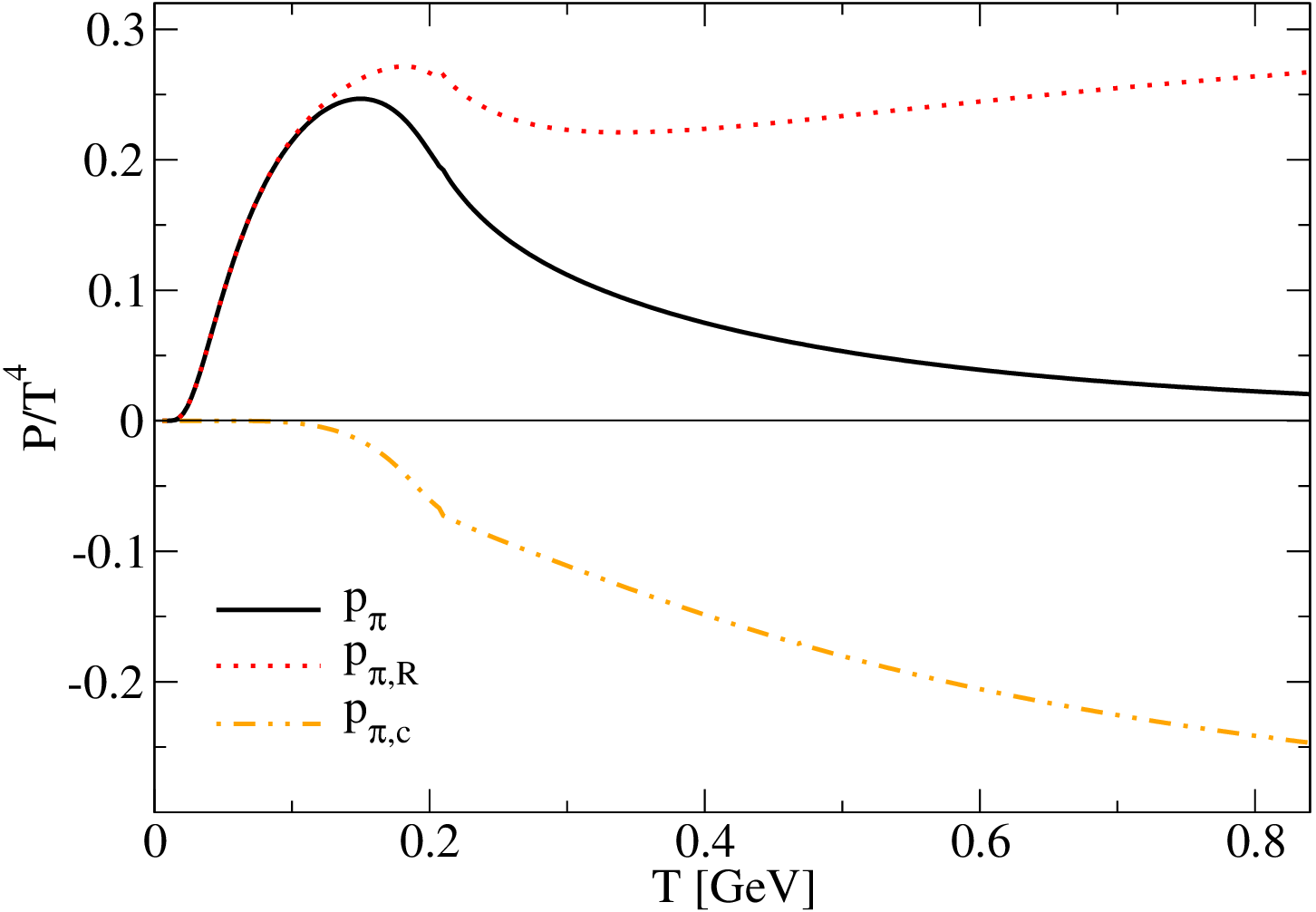}
}
\caption{
Right panel: 
Pionic contribution to the pressure, Eq.~(\ref{p_pi}), as a function of the 
temperature $T$ for chemical potential $\mu=0$ (solid line) and its 
decomposition into resonant (dotted line) and continuum (dash-double-dotted 
line) scattering contributions according to the decomposition of the phase 
shift in the density of states Eq.~(\ref{dos}). 
Left panel: Closeup of the right panel in the temperature region of the pion 
Mott transition at $T_{\pi,{\rm Mott}}=208$ MeV and comparison to different 
approximations for the resonance contribution: 
The dashed line represents the Breit-Wigner approximation (\ref{D_BW}) 
for the resonance contribution and shows the effect of the resonance broadening
compared to the pion bound state approximation (\ref{D_bound}), with a 
$T-$dependent mass but without broadening (dot-double-dashed line).
The dash-dotted line shows the total pion pressure when for the resonance 
contribution the Breit-Wigner approximation is used. 
Note that the effect of the continuum pressure sets in already before the Mott 
transition temperature is reached.
} 
\label{Fig:pionPressure}
\end{figure}

We want to give here a qualitative discussion of the physics content of the 
GBU EoS (\ref{GBU2}) by considering the phase shifts for pions at
rest in the medium, as shown in Fig.~\ref{Fig:Phi_matrix}, and suggesting 
their approximate boost invariance depending on $\omega$ and ${\bf q}$ only 
via the Mandelstam variable $s=\omega^2-q^2$ in the form
$\delta_\pi(\omega,{\bf q}=0)=\delta_\pi(\sqrt{s},{\bf q}=0)
\equiv\delta_\pi(s;T)$. 
Then, the GBU EoS for the pionic pressure can be given the suggestive 
form\footnote{Formally, when substituting the $\omega-$ integral in 
(\ref{GBU2}) for a $s-$ integral, the lower limit $\omega_{\rm min}=0$ 
corresponds to $s_{\rm min}=-q^2$. Because of the interpretation of the 
variable $s$ as a squared meson mass, we restrict the range of the $s-$ 
integration to all nonnegative values.}
\bea
\label{p_pi}
p_{\pi}(T)= \int_0^\infty ds ~D_\pi(s;T)~p_{\pi}(T,s)	~,
\eea
where 
\bea
\label{p_pi_s}
p_{\pi}(T,s)= -d_{\pi} T \int\frac{{\rm d}^3q}{(2\pi)^3}~
	\ln\left(1-\ee^{-\sqrt{q^2+s}/T}\right)	~
\eea
is the pressure of a relativistic Bose gas of pions with a ficticious mass 
$\sqrt{s}$.
The distribution of these masses to be integrated over for obtaining the 
total pressure is given by the density of states 
\bea
\label{dos}
D_\pi(s;T)=\frac{1}{\pi} \frac{d\delta_{\pi}(s;T)}{ds}~.
\eea
Exploiting the analytic decomposition of the total phase shift (\ref{delta})
into a continuum and a resonant contribution, 
$\delta_\pi=\delta_{\pi,c}+\delta_{\pi,R}$, 
we can separate the pion pressure into a negative continuum contribution and 
a resonance contribution, 
$p_{\pi}(T)=p_{\pi,c}(T)+p_{\pi,R}(T)$, 
where for the latter we can make use of the Breit-Wigner approximation 
(\ref{B-W})
\bea
\label{D_BW}
D_{\pi,R}^{BW}(s;T)=\frac{1}{\pi}\frac{\Gamma_\pi m_\pi}
{(s-m_\pi^2)^2+\Gamma_\pi^2 m_\pi^2}~,
\eea
for which holds that 
\bea
\label{D_bound}
\lim_{\Gamma_\pi \to 0} D_{\pi,R}^{BW}(s;T) = D_{\pi,{\rm bound}}=
\delta(s-m_\pi^2)~.
\eea
In the limit of vanishing  width $\Gamma_\pi\to 0$ the resonance goes over to 
an ideal bound state with infinite lifetime. 
Then the $s-$ integration in (\ref{p_pi}) can be analytically 
performed and yields the ideal pion gas form $p_{\pi}(T,m_\pi^2)$, here with a 
temperature dependent mass $m_\pi=m_\pi(T)$. 
Since the pion mass is rising with temperature 
(see Fig.~\ref{fig:massspectra}), in particular for $T>T_{X, {\rm Mott}}$, 
the corresponding pressure shows a slight drop (dot-double-dashed line in  
Fig.~\ref{Fig:pionPressure}).
 
Taking into account additionally the rapidly growing pion width 
$\Gamma_\pi(T)$ with a sharp onset at $T=T_{X, {\rm Mott}}$ 
(see Fig.~\ref{fig:massspectra}), the pion pressure is reduced stronger above
the Mott temperature (dashed line in Fig.~\ref{Fig:pionPressure}).
In these approximations, the role of the continuum has been neglected.
The pressure of the continuum states $p_{\pi,c}(T)$ on the other hand is 
obtained when the resonance contribution 
(dotted line in Fig.~\ref{Fig:pionPressure})
to the density of states is neglected,
\ie 
\bea
\label{D_c} 
D_\pi(s;T)=D_{\pi,c}(s;T)=\frac{1}{\pi} \frac{d\delta_{\pi,c}(s;T)}{ds}
\eea
is used in Eq.~(\ref{p_pi}).  
The result is shown as the dash-double-dotted
line in Fig.~\ref{Fig:pionPressure}; its 
contribution is negative and sets in already for $T<T_{X, {\rm Mott}}$.
The total pion pressure (solid line in Fig.~\ref{Fig:pionPressure}) with 
contributions from both resonant and continuum scattering shows a typical 
behaviour with increasing temperature: 
First a rise towards the Stefan-Boltzmann limit which, however, is never 
reached because the lowering of the continuum edge due to the chiral 
phase transition induces a reduction of the meson gas pressure already before
the Mott temperature is reached.
Second, above the Mott temperature, the growing pion width leads to a 
stronger reduction of the pressure with a rather sharp onset of this effect.
The resulting pattern appears like a ``shark fin''.

\subsubsection{Mott dissociation of diquarks}

The introduction diquark fields is a prerequisite for the description of 
baryons as quark-diquark bound states in a chiral quark model.
Thereby the diquark appears not necessarily as a bound state. Actually, in 
particular in QCD DSE approaches the diquark is unbound while the baryon is
a bound state, like in a Borromean three-particle state.

\begin{figure}[htbp] 
\centering{
	\includegraphics[width=0.48\textwidth]{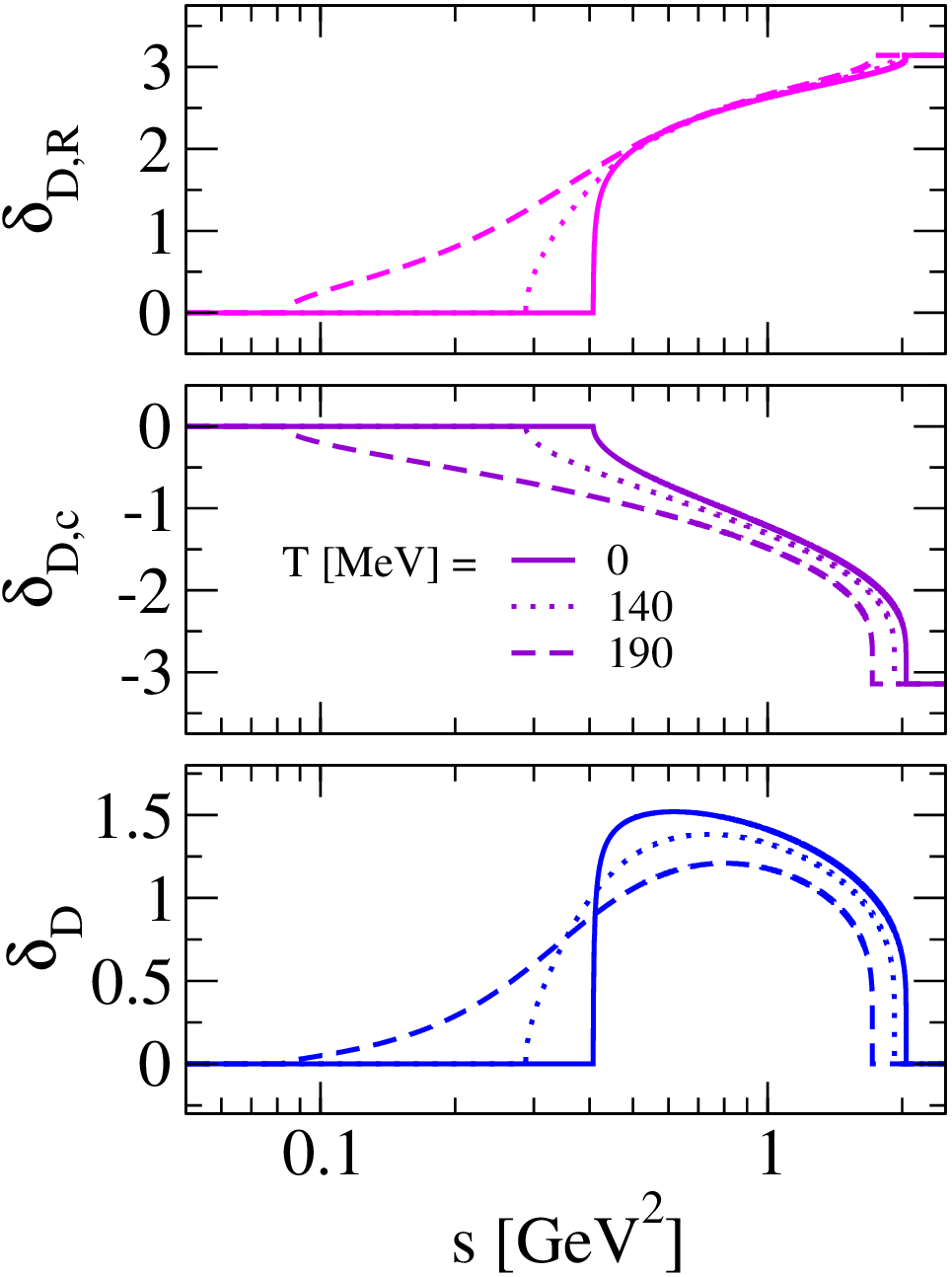}
	\includegraphics[width=0.48\textwidth]{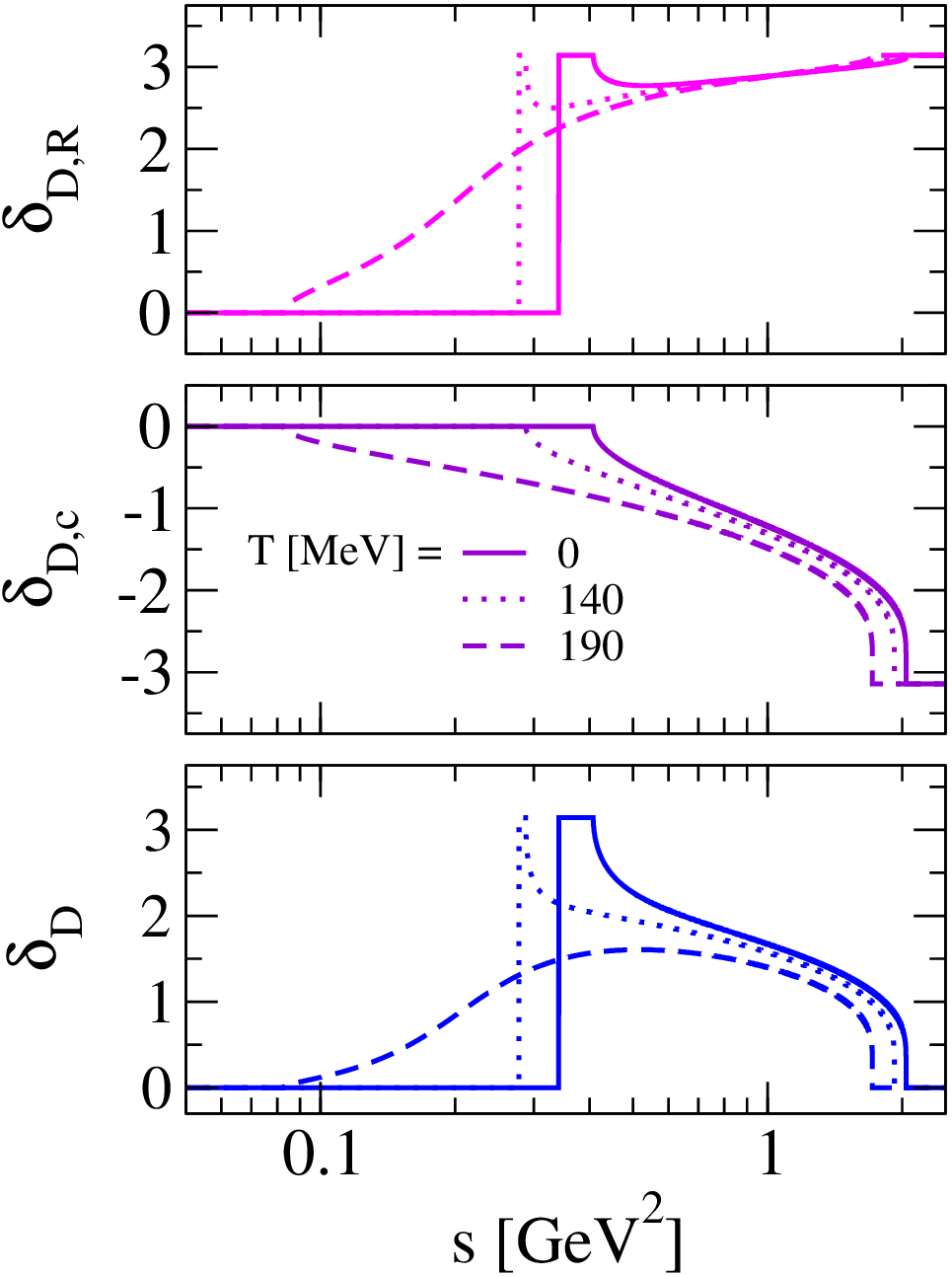}
}
\caption{
Phase shifts for scalar diquark channel with $G_D=3/4~G_S$ (left panels) 
and $G_D=G_S$ (right panels) at rest (${\bf q}=0$) as functions of the 
squared mass variable $s=\omega^2$ 
for different temperatures $T=0,~140,~190$ MeV. 
The upper panels show the resonant phase shifts which 
together with the continuum phase shifts shown in the middle panels add up
to the total phase shifts of the bottom panels.
Comparison to Fig.~\ref{Fig:Phi_matrix} shows that for moderate coupling 
(left panels) the diquark behaves ``sigma-like'', since it is unbound for all 
temperatures.
For strong coupling (right panels) the diquark behaves ``pion-like'', since
it is a bound state for low temperatures and exhibits a Mott transition at 
$T\approx 140$ MeV where the phase shift at the continuum threshold 
$s_{\rm thr}=4m^2$ jumps from the value $\pi$ to zero in accordance with the 
Levinson theorem (\ref{levinson}).
\label{Fig:Phi_D_matrix}}
\end{figure}

Having developed in the present work the theoretical basis for the description 
of the thermodynamics of mesons and diquarks in hot and dense quark matter
including a Mott dissociation transition, we want to discuss now after the 
pion also the diquark case. 
To that end we focus here on the normal quark matter phase and evaluate the 
diquark phase shifts for vanishing chemical potential at finite temperatures 
which encode the analytic properties of the diquark propagator.  
The results are shown in Fig.~\ref{Fig:Phi_D_matrix} for two cases of the 
diquark coupling strength: the moderate one  (left panels) which corresponds
to the Fierz value $G_D=3/4~G_S$ and the strong coupling case (right panels)
with $G_D=G_S$.
In the latter case the diquark is a bound state at zero temperature and 
its phase shift behaves similar to that of the pion in 
Fig.~\ref{Fig:Phi_matrix} which exhibits a Mott transition to an unbound 
resonance in the continuum at a certain temperature. 
For our parametrization this happens at $T_{D,{\rm Mott}}\approx 140$ MeV.
For the moderate coupling case, the diquark phase shift is shown in the left 
panels of Fig.~\ref{Fig:Phi_matrix} and behaves similar to the sigma meson:
it is already an unbound resonant scattering state in the vacuum at $T=0$ and 
becomes even less correlated when the temperature is increased. 

\begin{figure}[htbp] 
\centering{
	\includegraphics[width=0.48\textwidth]{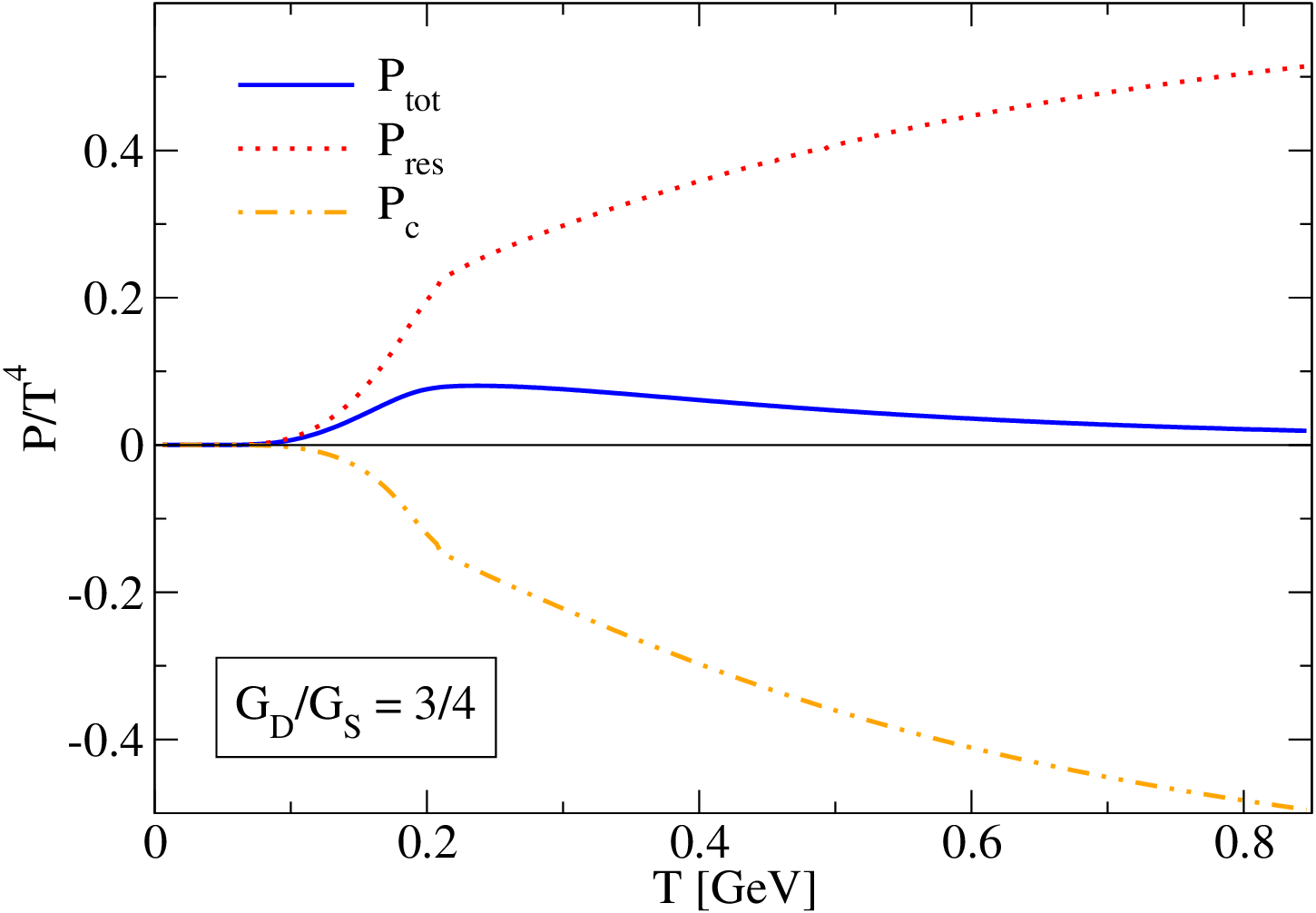}
	\includegraphics[width=0.48\textwidth]{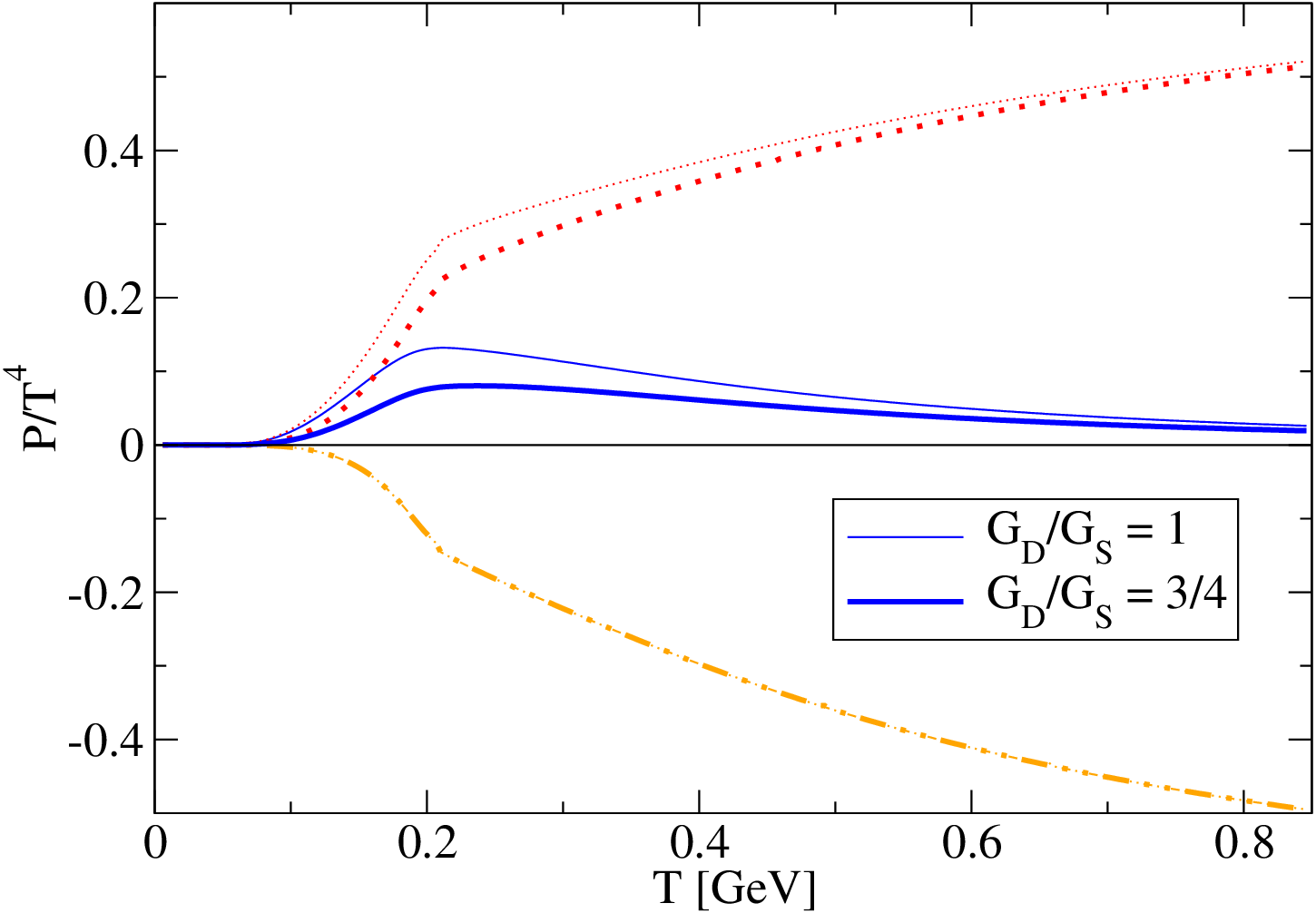}
}
\caption{
Pressure for the scalar diquark (solid line) with its resonant (dotted line)
and continuum (dash-double-dotted line) components as functions of temperature.
Bold lines are for the case of the moderate (Fierz) coupling $G_D/G_S=3/4$ 
(left and right panel) while thin lines are for the strong coupling case 
$G_D/G_S=1$ (right panel).  
}
\label{Fig:P_D_matrix}
\end{figure}

The diquark contribution to the thermodynamics is evaluated according to the 
Beth-Uhlenbeck formula (\ref{GBU-p}) for the pressure, where as in the pion 
case the ``no sea'' approximation is applied, \ie the contribution of the 
zero-point energy in (\ref{GBU}) is removed.
The result is shown in Fig.~\ref{Fig:P_D_matrix} for the Fierz coupling case 
(left panel) which exhibits an almost complete compensation between the 
contrinutions of the diquarks resonance and the scattering continuum so that 
the resulting pressure (solid line) in the diquark channel is much smaller 
than that of the pion channel albeit with a similar shape. 
The comparison with the strong coupling case (right panel) shows that the 
continuum background pressure is unaffected while the resonance contribution 
with the diquark Mott transition shows a moderate overall increase.  

\subsection{Further developments}

A key lesson from this investigation into the GBU EoS for the 
two-particle correlations in the pion and diquark channels in hot quark matter 
is that accounting for the broadening of hadronic states due to the Mott 
transition alone it is not sufficient and would violate the Levinson theorem  
\cite{Wergieluk:2012gd,Dubinin:2013yga}. 
One has to account for the scattering state continuum which influences the 
thermodynamics as soon as the thermal energy is of the order of the 
dissociation threshold, the gap between bound and continuum states of the 
spectrum. 
This appears to be an important modification of the so-called ``chemical'' 
picture where hadron species form a statistical ensemble of quasiparticles
which goes beyond attributing to these hadronic resonances a finite lifetime
(width). The more elaborate ``physical'' picture requires to account for 
a scattering state continuum which in a systematic way is given by a virial
expansion and as such appears in the GBU EoS. 
For a discussion of the cluster virial expansion in nuclear matter see, \eg
\cite{Ropke:2012qv}. 

For dense hadronic matter, where hadrons play the role of clusters, a similar
approach is still to be developed. We suggest that this could be done along 
the lines of the GBU approach given here.  
However, the problem one is facing when one wants to build such an approach 
on an NJL-type model is the lack of confinement.
In the NJL model used here, just the pions are bound states in the vacuum 
while the sigma meson is already an unbound resonance, decaying to quarks. 
In the flavor SU(3) extension of the NJL model, it is just the pseudoscalar 
meson octet which is bound while the other channels important for the 
phenomenology (vector, scalar and axialvector mesons) are unbound.

While a solution to this severe problem has still to be worked out, there
are simple extensions of the NJL model which can be used to remove unphysical
quark decay thresholds but still keep the microscopic description of the 
hadron Mott effect sufficiently simple. 
One of such extensions mimicks confining interactions by suppressing long 
wavelength modes in quark propagation by applying a low-momentum 
cutoff to quark loop integrals. 
Moreover, since confinement and chiral symmetry breaking shall be related, 
this cutoff has been fixed to the dynamically generated quark mass gap which 
vanishes upon chiral symmetry restoration. 
For details, see \cite{Dubinin:2013yga} and references therein. 
In this work it has been shown how this infrared cutoff increases the 
threshold for quark-antiquark scattering states and thus makes the $\sigma-$
meson a bound state in the chiral symmetry broken phase for low temperatures. 
It could also be demonstrated in \cite{Dubinin:2013yga} within this 
NJL model generalized by a low-momentum cutoff, that the shape of the 
phase shifts and their temperature dependence for the   $\sigma-$meson becomes 
similar to that of the pion, where the transition from a bound state to a 
resonance in the continuum can be recognized in accordance with the Levinson 
theorem.
Summarizing this section, we would like to point out that the NJL model and 
its modifications provide a useful tool for a field theoretic formulation of
the physics of the Mott dissociation of hadrons as quark bound states in hot,
dense quark matter or, conversely, of the hadronization of quark correlations 
in an expanding and cooling QGP.

\section{Conclusions}
\label{sec:conclusions}

In this work, we have presented a general approach to the discussion of 
two-particle correlations in quark matter within a field theoretical chiral 
quark model of the NJL type.
By restricting the expansion of the fermion determinant of the bosonized 
partition function to the  gaussian approximation, the path integral over the
meson and diquark fluctuations can be performed and a closed expression is 
obtained in the form of the determinant of a generalized matrix propagator
for two-particle states (meson and diquark fields).
The diagonalization of this matrix defines the mode spectrum of the model. 
A detailed discussion of limiting cases (chiral limit, zero momentum limit) 
is performed and analytic expressions are given.

Nambu-Goldstone (NG) modes in the 2SC phase are considered and the 
finite $T$ extension of the work by Ebert and collaborators \cite{Ebert:2004dr}
is provided here for the first time.
It is demonstrated that in the 2SC phase there are only 3 instead of 5
massless NG modes, in accordance with the Nielsen-Chadha theorem. 

We have discussed the interplay between bound and scattering states in the 
medium, in particular at the Mott transition where the bound state transforms 
to a resonance in the continuum. 
This transition is reflected in a vanishing of the binding energy as well as a 
jump by $\pi$ of the phase shift at threshold in acordance with the Levinson 
theorem.

The thermodynamic potential is given the form of a generalized Beth-Uhlenbeck 
equation where special emphasis is on the discussion of the role of 
the continuum contributions as well as of the resonant two-particle 
correlations as expressed in terms of the in-medium scattering phase shifts. 
It is shown that accounting for the spectral broadening of hadrons alone is 
insufficient to account for the Mott dissociation. 
Only together with the continuum states can be garanteed that the result for
the pressure is in accordance with the in-medium Levinson theorem.

The in-medium phase shifts for mesonic and diquark correlation channels as 
evaluated in this work within a NJL model description of low energy QCD 
contain spectral information which may be exploited to study mesonic and 
diquark correlation functions at finite temperature, around the chiral 
restoration.
It will be very interesting to investigate how the Mott transition for mesons
manifests itself in these functions for which also ab-initio calcuations using 
lattice QCD simulations exist. We plan to extend our work in this direction
and to provide guidance from this low-energy QCD model for the interpretation
of those lattice QCD data.  

An outlook is given to the further development of the approach towards a
description of meson-baryon matter on the 
basis of chiral quark models of the NJL type which shall be detailed in 
subsequent work.

\section*{Acknowledgement}

We acknowledge R.~Anglani for his collaboration in an early stage of this work
and T.~Brauner for discussions and his critical reading of the manuscript.
This work was supported by the Deutsche Forschungsgemeinschaft (DFG) under 
contract BU 2406/1-1 and by the Polish National Science Centre 
within the ``Maestro'' programme under grant no. UMO-2011/02/A/ST2/00306.
The work of D.B. was supported in part by the Polish Ministry of Science and 
Higher Education (MNiSW) under grant no. 1009/S/IFT/14. 
G.R. and D.B. are grateful to the Erasmus programme and partnership 
between Universities of Rostock and Wroclaw which supported mutual visits.  

\appendix
\renewcommand*{\thesection}{\Alph{section}}
\section{Meson and diquark polarization functions}
\label{App:Integrals}

In this appendix we give explicit expressions for the elements of the 
polarization matrix $\Pi$ defined in Eq.~(\ref{mixedPropagator}).
We thereby neglect the vector fields $\omega_\mu$. 
Moreover, since the three isospin components of the pion are degenerate and do 
not mix with each other or any other mode, we only list one generic component 
here for simplicity.

\subsection{2SC phase}

In the 2SC phase,  the polarization matrix has the structure
\bea
\label{App:polarizationMatrix}
\Pi
=
\begin{pmatrix}
\Pi_{\pi\pi}&0&0&0&0&0&0&0\\
0&\Pi_{\sigma\sigma}&\Pi_{\sigma\delta_2}&\Pi_{\sigma\delta_2^*}&0&0&0&0\\
0&\Pi_{\delta_2^*\sigma}&\Pi_{\delta_2^*\delta_2}&\Pi_{\delta_2^*\delta_2^*}&0&0&0&0\\
0&\Pi_{\delta_2\sigma}&\Pi_{\delta_2\delta_2}&\Pi_{\delta_2\delta_2^*}&0&0&0&0\\
0&0&0&0&\Pi_{\delta_5^*\delta_5}&0&0&0\\
0&0&0&0&0&\Pi_{\delta_5\delta_5^*}&0&0\\
0&0&0&0&0&0&\Pi_{\delta_7^*\delta_7}&0\\
0&0&0&0&0&0&0&\Pi_{\delta_7\delta_7^*}
\end{pmatrix}
~.
\eea

Each matrix element depends on an external three-momentum $\bf q$
and an external bosonic Matsubara frequency $iz_n$, which 
after analytic continuation becomes the complex variable $z$.
Labelling the fields in the vector (\ref{X}) by 
$i=\vec{\pi}^T,\sigma,\delta^*_2,\delta_2,\delta^*_5,\delta_5,\delta^*_7,\delta_7$ and 
$j=\vec{\pi},\sigma,\delta_2,\delta^*_2,\delta_5,\delta^*_5,\delta_7,\delta^*_7$, we have
\bea
         \Pi_{\rm ij}  
         \equiv 
         \Pi_{\rm ij} (\ii z_n,{\bf q})
         \equiv 
         \Pi_{\rm ij} (z,{\bf q})~.
\eea
The explicit expressions aslisted below also contain an internal momentum 
$\bf p$ which is integrated over. 
It is then convenient to define a third momentum 
${\bf k} = {\bf p} - {\bf q}$. 

The individual elements are as follows
\bea
\Pi_{\pi\pi}
&=&
\frac{1}{2}
N_f\int\frac{{\rm d}^3p}{(2\pi)^3}~ \sum_{s_p,s_k}
\mathcal{T}_-^+(s_p,s_k)
\bigg\{
\frac{n(s_p\xi_{\bf p}^{s_p})-n(s_k\xi_{\bf k}^{s_k})}
       {z-s_k\xi_{\bf k}^{s_k}+s_p\xi_{\bf p}^{s_p}}
-
\frac{n(s_p\xi_{\bf p}^{s_p})-n(s_k\xi_{\bf k}^{s_k})}
       {z+s_k\xi_{\bf k}^{s_k}-s_p\xi_{\bf p}^{s_p}}
\nonumber \\&&
+\sum_{t_p,t_k}
F(s_p,s_k;t_p,t_k)
\left(
t_pt_kE_{\bf p}^{s_p}E_{\bf k}^{s_k} + s_ps_k\xi_{\bf p}^{s_p}\xi_{\bf k}^{s_k} - |\Delta_{\rm MF}|^2
\right)
\bigg\}\\
\Pi_{\delta_2\delta_2}
&=&
\frac{1}{4}
(\Delta_{\rm MF}^*)^2N_f
\int\frac{{\rm d}^3p}{(2\pi)^3}~
\sum_{\substack{s_p,t_p\\s_k,t_k}}
\mathcal{T}_+^+(s_p,s_k)
F(s_p,s_k;t_p,t_k)
\\
\Pi_{\delta_2^*\delta_2}
&=&
\frac{1}{4}
N_f
\int\frac{{\rm d}^3p}{(2\pi)^3}~
\sum_{\substack{s_p,t_p\\s_k,t_k}}
\mathcal{T}_+^+(s_p,s_k)
F(s_p,s_k;t_p,t_k)
(t_pE_{\bf p}^{s_p} + s_p\xi_{\bf p}^{s_p})
(t_kE_{\bf k}^{s_k} - s_k\xi_{\bf k}^{s_k})
\\
\Pi_{\sigma\delta_2}
&=&
\frac{1}{4}
m\Delta_{\rm MF}^*N_f
\int\frac{{\rm d}^3p}{(2\pi)^3}~\sum_{\substack{s_p,t_p\\s_k,t_k}}
\left(
	\frac{s_p}{E_{\bf p}} + \frac{s_k}{E_{\bf k}}
\right)
F(s_p,s_k;t_p,t_k)
\left[
\left(t_pE_{\bf p}^{s_p} - t_kE_{\bf k}^{s_k}\right) 
+ \left(s_p\xi_{\bf p}^{s_p} + s_k\xi_{\bf k}^{s_k}\right)
\right]
\\
\Pi_{\delta_5^*\delta_5}
&=&
\frac{1}{2}
N_f
\int\frac{{\rm d}^3p}{(2\pi)^3}~
\sum_{\substack{s_p,t_p\\s_k,t_k}}
\mathcal{T}_+^+(s_p,s_k)
\left\{
	\frac{E_{\bf p}^{s_p}-s_pt_p\xi_{\bf p}^{s_p}}{E_{\bf p}^{s_p}}
	\frac{n(s_k\xi_{\bf k}^{s_k})-n(t_pE_{\bf p}^{s_p})}
	{z-t_pE_{\bf p}^{s_p}+s_k\xi_{\bf k}^{s_k}}
	+
	\frac{E_{\bf k}^{s_k}-s_kt_k\xi_{\bf k}^{s_k}}{E_{\bf k}^{s_k}}
	\frac{n(s_p\xi_{\bf p}^{s_p})-n(t_kE_{\bf k}^{s_k})}
	{z-t_kE_{\bf k}^{s_k}+s_p\xi_{\bf p}^{s_p}}
\right\}
	\nonumber\\
\eea
and the remaining elements are recast by the replacements 
\bea
\label{App:sym0}
\Pi_{\sigma\sigma}&=&\Pi_{\pi\pi}\left(\mathcal{T}_-^+\to\mathcal{T}_-^-\right)
\\ 
\label{App:sym1}
\Pi_{\delta_2^*\delta_2^*}&=&\Pi_{\delta_2\delta_2}
\left((\Delta_{\rm MF}^*)^2\to\Delta_{\rm MF}^2\right)
\\
\label{App:sym2}
\Pi_{\delta_2\delta_2^*}&=&\Pi_{\delta_2^*\delta_2}
\left(z\to-z,{\bf p}\leftrightarrow{\bf k}\right)
\\ 
\label{App:sym3}
\Pi_{\sigma\delta_2^*}&=&\Pi_{\sigma\delta_2}
\left(\Delta_{\rm MF}^*\to\Delta_{\rm MF},{\bf p}\leftrightarrow{\bf k}\right)
\\
\label{App:sym4}
\Pi_{\delta_2^*\sigma}&=&\Pi_{\sigma\delta_2}
\left(\Delta_{\rm MF}^*\to\Delta_{\rm MF},z\to-z\right)
\\ 
\label{App:sym5}
\Pi_{\delta_2\sigma}&=&\Pi_{\sigma\delta_2^*}
\left(\Delta_{\rm MF}\to\Delta_{\rm MF}^*,z\to-z\right)
\\
\label{App:sym6}
\Pi_{\delta_5\delta_5^*}&=&\Pi_{\delta_5^*\delta_5}
\left(z\to-z,{\bf p}\leftrightarrow{\bf k}\right)
\\ 
\label{App:sym7}
\Pi_{\delta_7\delta_7^*}&=&\Pi_{\delta_5\delta_5^*}
\\ 
\label{App:sym8}
\Pi_{\delta_7^*\delta_7}&=&\Pi_{\delta_5^*\delta_5}
\eea
Here $s_p,s_k;t_p,t_k = \pm 1$ are sign operators, 
$E_{\bf p}$ , $E_{\bf p}^\pm$,  and $\xi_{\bf p}^\pm$ are the single-quark 
dispersion relations, given in Sect.~\ref{secMeanField}, and $n(x)$ denotes 
again the Fermi distribution function.
Moreover, we have defined
\bea
\label{F-kernel}
	F(s_p,s_k;t_p,t_k)
	=
	\frac{t_pt_k}{E_{\bf p}^{s_p}E_{\bf k}^{s_k}}
	\frac{n(t_pE_{\bf p}^{s_p}) - n(t_kE_{\bf k}^{s_k})}
       {z - t_kE_{\bf k}^{s_k} + t_pE_{\bf p}^{s_p}}~,
\eea
and the kinematic pre-factors
\bea
	\mathcal{T}_\mp^\pm(s,s')
	=
	1\mp ss'\frac{{\bf p}\cdot{\bf k}\pm m^2}{E_{\bf p}E_{\bf k}}~.
\eea

\subsection{Correlations at rest}
\label{App:IntegralsRestFrame}

The expressions for the polarization functions get strongly simplified for
correlations at rest, ${\bf q} = 0$.
We then have ${\bf p}={\bf k}$, which then puts restrictions on the summation 
over $s_p$ and $s_k$ via $\mathcal{T}_\mp^\pm$ and at last we can carry out the
summations over $t_p,t_k$.
This then leaves us with only one sign operator, which is then associated with 
particle ($-$) and anti-particle ($+$) contributions.
We find
\bea
	\Pi_{\pi\pi}(z,{\bf 0})
	&=&
	-8I_\pi(z)
	~,
	\\
	\Pi_{\sigma\sigma}(z,{\bf 0}) 
	&=&
	-8I_\sigma(z)-16m^2|\Delta_{\rm MF}|^2I_4(z)
	~,
	\\
	\Pi_{\delta_2^*\delta_2}(z,{\bf 0})
	&=&
	2I_\Delta+4zI_1(z)+(4|\Delta_{\rm MF}|^2-2z^2)I_0(z)
	~,
	\\
	\Pi_{\delta_2\delta_2}(z,{\bf 0})
	&=&
	4\Delta_{\rm MF}^2I_0(z)
	~,
	\\
	\Pi_{\sigma\delta_2}(z,{\bf 0})
	&=&
	-4m\Delta_{\rm MF}(zI_2(z)+2I_3(z))
	~,
	\\
	\Pi_{\delta_5^*\delta_5}(z,{\bf 0})
	&=&
	I_\Delta+2zI_7(z)-(|\Delta_{\rm MF}|^2-z^2)I_5(z)
	~,
\eea
and the remaining elements can be found by applying the symmetry relations Eqs.~(\ref{App:sym1})-(\ref{App:sym8}).
The constant $I_\Delta$ and the functions $I_i(z)$ are defined as follows:
\bea
	I_\Delta=I_\Delta^++I_\Delta^- &~,& I_0=I_0^++I_0^-~,\\
	I_1=I_1^+-I_1^- &~,& I_2=I_2^++I_2^-~,\\
	I_3=I_3^++I_3^- &~,& I_4=I_4^++I_4^-~,\\
	I_5=I_5^++I_5^- &~,& I_7=I_7^+-I_7^-~,
\eea
where the individual terms are
\bea
	I_\Delta^\pm
	\equiv
	\int\frac{{\rm d}^3p}{(2\pi)^3}~
	\left[1-2n(E_{\bf p}^\pm)\right]\frac{1}{E_{\bf p}^\pm,}
	&~,&	
	I_0^\pm(z)
	\equiv
	\int\frac{{\rm d}^3p}{(2\pi)^3}~
	\frac{1}{E_{\bf p}^\pm}F_{\bf p}^\pm(z)
	~,\\
	I_1^\pm(z)
	\equiv
	\int\frac{{\rm d}^3p}{(2\pi)^3}~
	\frac{\xi_{\bf p}^\pm}{E_{\bf p}^\pm}F_{\bf p}^\pm(z)
	&~,&	
	I_2^\pm(z)
	\equiv
	\int\frac{{\rm d}^3p}{(2\pi)^3}~
	\frac{1}{E_{\bf p}}\frac{1}{E_{\bf p}^\pm}F_{\bf p}^\pm(z)
	~,\\
	I_3^\pm(z)
	\equiv
	\int\frac{{\rm d}^3p}{(2\pi)^3}~
	\frac{1}{E_{\bf p}}\frac{\xi_{\bf p}^\pm}{E_{\bf p}^\pm}F_{\bf p}^\pm(z)
	&~,&	
	I_4^\pm(z)
	\equiv
	\int\frac{{\rm d}^3p}{(2\pi)^3}~
	\frac{1}{E_{\bf p}^2}
	\frac{1}{E_{\bf p}^\pm}
	F_{\bf p}^\pm(z)
	\\
	I_5^\pm(z)
	\equiv
	\int\frac{{\rm d}^3p}{(2\pi)^3}~
	\frac{1-2n(E_{\bf p}^\pm)}
	{E_{\bf p}^\pm(z^2\pm z\xi_{\bf p}^\pm-|\Delta_{\rm MF}|^2)}
	&~,&	
	I_7^\pm(z)
	\equiv
	\int\frac{{\rm d}^3p}{(2\pi)^3}~
	\frac{1-2n(\xi_{\bf p}^\pm)}
	{z^2\pm z\xi_{\bf p}^\pm-|\Delta_{\rm MF}|^2}
\eea
and the remaining integral for the mesons is given by
\bea
\label{Ipi}
	I_\pi(z)
	&\equiv&
	\int\frac{{\rm d}^3p}{(2\pi)^3}~
	\left\{
		\left[1-n(\xi_{\bf p}^+)-n(\xi_{\bf p}^-)\right]
		\frac{2E_{\bf p}}{z^2-4E_{\bf p}^2}
		\right.\nonumber\\
		&&
		+
		\frac{E_{\bf p}^+E_{\bf p}^--\xi_{\bf p}^+\xi_{\bf p}^--|\Delta_{\rm MF}|^2}{E_{\bf p}^+E_{\bf p}^-}
		\left[n(E_{\bf p}^-)-n(E_{\bf p}^+)\right]
		\frac{E_{\bf p}^+-E_{\bf p}^-}{z^2-\left(E_{\bf p}^+-E_{\bf p}^-\right)^2}
		\nonumber\\
		&&\left.
		+
		\frac{E_{\bf p}^+E_{\bf p}^-+\xi_{\bf p}^+\xi_{\bf p}^-+|\Delta_{\rm MF}|^2}{E_{\bf p}^+E_{\bf p}^-}
		\left[1-n(E_{\bf p}^+)-n(E_{\bf p}^-)\right]
		\frac{E_{\bf p}^++E_{\bf p}^-}{z^2-\left(E_{\bf p}^++E_{\bf p}^-\right)^2}
	\right\}
	~.
\eea
The integral $I_\sigma(z)$ differs from $I_\pi(z)$ only by an additional 
overall factor ${\bf p}^2/E_{\bf p}^2$ in the integrand.
\\
In the normal phase, for $\Delta_{\rm MF}=0$, the integral (\ref{Ipi})
reduces to the simple form
\bea
\label{Ipi0}
	I_{\pi,0}(z)
	&=&
	N_c\int\frac{{\rm d}^3p}{(2\pi)^3}~
		\left[1-n(\xi_{\bf p}^+)-n(\xi_{\bf p}^-)\right]
		\frac{2E_{\bf p}}{z^2-4E_{\bf p}^2}~.
\eea
The function 
\bea
	F_{\bf p}^\pm(z)
	\equiv
	\frac{1-2n(E_{\bf p}^\pm)}{z^2-4\left(E_{\bf p}^\pm\right)^2}
\eea
is the analogue to the function $F(s_p,s_k;t_p,t_k)$ defined above in 
(\ref{F-kernel}) for correlations at rest (${\bf q}=0$).

\subsection{Mesonic polarization functions in the normal phase}
\label{App:PolarizationNormal}

In the normal phase the meson and diquark-modes decouple. 
We denote the mesonic matrix elements by
$\Pi_\pi \equiv \Pi_{\pi\pi}$, 
$ \Pi_\sigma \equiv \Pi_{\sigma\sigma}$. 
They are explicitly given by the expression
\bea
\label{polMeson}
	\Pi_{\pi/\sigma}(z,{\bf q})
	&=&
\frac{1}{2}
	 N_fN_c
	\sum_{s_p,s_k}
	\int\frac{{\rm d}^3p}{(2\pi)^3}~
	\mathcal{T}_-^\pm(s_p,s_k)
	\frac{n^-(s_pE_{\bf p})-n^-(s_kE_{\bf k})}
	{z + s_pE_{\bf p} - s_kE_{\bf k}}
	+(\mu^*\to-\mu^*)~,
\eea
where $\mathcal{T}_-^+(s_p,s_k)$ holds for the pion case while 
$\mathcal{T}_-^-(s_p,s_k)$ for the sigma meson.
After analytic continuation to the complex $z$ plane, the analytic properties
of the polarization function are captured in its spectral density as defined 
by the imaginary part of the retarded function, 
$\IM\Pi_{\pi/\sigma}(\omega+\ii\eta,{\bf q})$.

Results for scalar and pseudo-scalar mesons have already been obtained by \eg 
\cite{Hatsuda:1994pi,Rossner:2007ik} and shall be summarized here.
\bea
	\IM
	\Pi_{\rm \sigma,\pi}(\omega+\ii\eta,{\bf q})
	=
	N_{\pi,\sigma}\frac{N_fN_c}{16\pi|{\bf q}|}
	\left\{
		\Theta(s-4m^2)
		\left[
			\Theta(\omega)J_{\rm M, pair}^+
			+
			\Theta(-\omega)J_{\rm M, pair}^-
		\right]
		+
		\Theta(-s)J_{\rm M, Landau}
	\right\}
\eea
with
\bea
	J_{\rm M, pair}^\pm
	&=&
	T\ln\left[
		\frac{[1-n^\pm(\mathcal{E}^-)]n^\mp(\mathcal{E}^-)}
			 {[1-n^\pm(\mathcal{E}^+)]n^\mp(\mathcal{E}^+)}
	\right]
	=
	T\ln\left[
		\frac{n^\mp(-\mathcal{E}^-)n^\mp(\mathcal{E}^-)}
			 {n^\mp(-\mathcal{E}^+)n^\mp(\mathcal{E}^+)}
	\right]
	\\
	J_{\rm M, Landau}
	&=&
	T\ln\left[
		\frac{n^+(\mathcal{E}^-)n^-(\mathcal{E}^-)}
			 {n^+(-\mathcal{E}^+)n^-(-\mathcal{E}^+)}
	\right]
	=		
	-2\omega
	+
	T\ln\left[
		\frac{n^+(-\mathcal{E}^-)n^-(-\mathcal{E}^-)}
			 {n^+(\mathcal{E}^+)n^-(\mathcal{E}^+)}	
	\right]
\eea
with the Fermi distribution functions 
$n^\pm(E)=[\exp\{(E\pm\mu^*)/T\}+1]^{-1}$ and
$
\mathcal{E}^\pm
=
\frac{\omega}{2}\pm\frac{|{\bf q}|}{2}\sqrt{1-\frac{4m^2}{s}}
$
.

The kinematic prefactors are $N_\pi=s$ and $N_\sigma=s-4m^2$, with 
$s=\omega^2-|{\bf q}|^2$.
We want to point out, that the imaginary part is an odd function of $\omega$ 
and consequently the real part is even in $\omega$ so that the spectral 
function is an odd function of $\omega$.

This form has the thresholds for the occurrence of imaginary parts, given 
in terms of $\Theta$-functions, explicitly; a property that has been exploited 
in the discussion of the Mott effect for mesons in Sect.~\ref{sec:GBU}.

\subsection{Diquark polarization functions in the normal phase}	
\label{App:PolarizationNormal-D}

In the normal phase all diquark modes are degenerate and we denote them as
$\Pi_{\rm D} \equiv \Pi_{\delta_A^* \delta_A}$ 
and 
$\Pi_{\rm \bar D} \equiv \Pi_{\delta_A \delta_A^*}$.
The diquark polarization function takes the form
\bea
	\label{polDiquark}
	\Pi_{\rm D}(z-2\mu^*,{\bf q})
	&=&
	 N_f 
	\sum_{s_p,s_k}
	\int\frac{{\rm d}^3p}{(2\pi)^3}~
	\mathcal{T}_-^+(s_p,s_k)
	\frac{n^+(s_pE_{\bf p})-n^-(s_kE_{\bf k})}
	{z + s_pE_{\bf p} - s_kE_{\bf k}}
	~.
\eea
The anti-diquark mode is then obtained by the symmetry relation 
$\Pi_{\bar{\rm D}}=\Pi_{\rm D}(\mu^*\to-\mu^*)$.

The imaginary part of the retarded diquark polarization function is evaluated 
to be
\bea
	\IM\Pi_{\rm D}(\omega+\ii\eta,{\bf q})
	&=&
	\frac{N_f}{16\pi|\bf q|}
	\sum_{a=\pm}
	s_a
	\left[
	\Theta(s_a-4m^2) J_{\rm D, pair}^a
	+
	\Theta(-s_a) J_{\rm D, Landau}^a
	\right]
\eea
with
\bea
	J_{\rm D, pair}^\pm
	&=&
	T\ln\left[
		\frac{1-n^\mp(\mathcal{E}_\pm^-)}{1-n^\mp(\mathcal{E}_\pm^+)}
		\frac{n^\mp(\mathcal{E}_\pm^-)}{n^\mp(\mathcal{E}_\pm^+)}
	\right]
	=
	T\ln\left[
		\frac{n^\pm(-\mathcal{E}_\pm^-)}{n^\pm(-\mathcal{E}_\pm^+)}
		\frac{n^\mp(\mathcal{E}_\pm^-)}{n^\mp(\mathcal{E}_\pm^+)}
	\right]
	~,
	\\
	J_{\rm D, Landau}^\pm
	&=&
	2T\ln\left[
		\frac{n^\mp(\mathcal{E}_\pm^-)}{n^\pm(-\mathcal{E}_\pm^+)}
	\right]
	=
	-2\omega
	+
	2
	T\ln\left[
		\frac{n^\pm(-\mathcal{E}_\pm^-)}{n^\mp(\mathcal{E}_\pm^+)}
	\right]
\eea
where we defined
$
\mathcal{E}_\mp^\pm
=
\frac{\omega_\mp}{2}\pm\frac{|{\bf q}|}{2}\sqrt{1-\frac{4m^2}{s_\mp}}
$
and $\omega_\pm=\omega\pm2\mu^*$ and $s_\pm=\omega_\pm^2-|{\bf q}|^2$.
Again, the real part of the spectral function is to be evaluated utilizing a 
Kramers-Kronig relation.
This result collapses to the pion imaginary part at $\mu^*=0$ up to the 
prefactor $2$.
For the case of degenerate flavors considered here, the Landau damping term
vanishes for correlations at rest (${\bf q} = 0$).

\newpage



\begin{thebibliography}{99}

\bibitem{Roberts:1994dr} 
  C.~D.~Roberts and A.~G.~Williams,
  Prog.\ Part.\ Nucl.\ Phys.\  {\bf 33}, 477 (1994).

\bibitem{Tandy:1997qf} 
  P.~C.~Tandy,
  Prog.\ Part.\ Nucl.\ Phys.\  {\bf 39}, 117 (1997).

\bibitem{Alkofer:2000wg} 
  R.~Alkofer and L.~von Smekal,
  Phys.\ Rept.\  {\bf 353}, 281 (2001).

\bibitem{Roberts:2000aa} 
  C.~D.~Roberts and S.~M.~Schmidt,
  Prog.\ Part.\ Nucl.\ Phys.\  {\bf 45}, S1 (2000).

\bibitem{Kleinert:1976xz} 
  H.~Kleinert,
  Phys.\ Lett.\ B {\bf 62}, 429 (1976).

\bibitem{Ebert:1976rh} 
  D.~Ebert and V.~N.~Pervushin,
  In *Tbilisi 1976, Proceedings, Conference On High Energy Physics, Vol.I*, Dubna 1976, C125-127

\bibitem{Cahill:1988zi}
  R.~T.~Cahill,
  Austral.\ J.\ Phys.\  {\bf 42}, 171 (1989).

\bibitem{Reinhardt:1989rw}
  H.~Reinhardt,
  Phys.\ Lett.\  {\bf B244}, 316-326 (1990).

\bibitem{Ebert:1994mf} 
  D.~Ebert, H.~Reinhardt and M.~K.~Volkov,
  Prog.\ Part.\ Nucl.\ Phys.\  {\bf 33}, 1 (1994).

\bibitem{Cahill:1988bh}
  R.~T.~Cahill, J.~Praschifka, C.~Burden,
  Austral.\ J.\ Phys.\  {\bf 42}, 161 (1989).

\bibitem{Burden:1988dt}
  C.~J.~Burden, R.~T.~Cahill, J.~Praschifka,
  Austral.\ J.\ Phys.\  {\bf 42}, 147 (1989).

\bibitem{Praschifka:1986nf}
  J.~Praschifka, C.~D.~Roberts, R.~T.~Cahill,
  Phys.\ Rev.\  {\bf D36}, 209 (1987).

\bibitem{Zuckert:1996nu} 
  U.~Zuckert, R.~Alkofer, H.~Weigel and H.~Reinhardt,
  Phys.\ Rev.\ C {\bf 55}, 2030 (1997).


\bibitem{Ebert:1985kz} 
  D.~Ebert and H.~Reinhardt,
  Nucl.\ Phys.\ B {\bf 271}, 188 (1986).


\bibitem{Christov:1995vm} 
  C.~V.~Christov, A.~Blotz, H.~-C.~Kim, P.~Pobylitsa, T.~Watabe, T.~Meissner, E.~Ruiz Arriola and K.~Goeke,
  Prog.\ Part.\ Nucl.\ Phys.\  {\bf 37}, 91 (1996).

\bibitem{Golli:1998rf} 
  B.~Golli, W.~Broniowski and G.~Ripka,
  Phys.\ Lett.\ B {\bf 437}, 24 (1998).

\bibitem{Alkofer:1994ph} 
  R.~Alkofer, H.~Reinhardt and H.~Weigel,
  Phys.\ Rept.\  {\bf 265}, 139 (1996).

\bibitem{Alkofer:1995mv} 
  R.~Alkofer and H.~Reinhardt,
  {\it Chiral quark dynamics}
  (Springer, Berlin, 1995).

\bibitem{Ripka:1997zb} 
  G.~Ripka,
 {\it Quarks bound by chiral fields} 
 (Clarendon Press, Oxford, 1997).

\bibitem{Sun:2007fc}
  G.~-f.~Sun, L.~He, P.~Zhuang,
  Phys.\ Rev.\  {\bf D75}, 096004 (2007).

\bibitem{Hufner:1994ma}
  J.~H\"ufner, S.~P.~Klevansky, P.~Zhuang, H.~Voss,
  Annals Phys.\  {\bf 234}, 225-244 (1994).

\bibitem{Wergieluk:2012gd} 
  A.~Wergieluk, D.~Blaschke, Y.~L.~Kalinovsky and A.~Friesen,
  Phys. Part. Nucl. Lett. {\bf 10}, 660 (2013).

\bibitem{Yamazaki:2012ux} 
  K.~Yamazaki and T.~Matsui,
  Nucl.\ Phys.\ A {\bf 913}, 19 (2013).

\bibitem{Dubinin:2013yga} 
  A.~Dubinin, D.~Blaschke and Y.~L.~Kalinovsky,
  Acta Phys. Pol. B Proc. Suppl. {\bf 7}, 215 (2014).

\bibitem{Rossner:2007ik} 
  S.~Roessner, T.~Hell, C.~Ratti and W.~Weise,
  Nucl.\ Phys.\ A {\bf 814}, 118 (2008).

\bibitem{Wang:2010iu}
  J.~-C.~Wang, Q.~Wang, D.~H.~Rischke,
  Phys.\ Lett.\  {\bf B704}, 347 (2011).

\bibitem{Lawley:2006ps}
  S.~Lawley, W.~Bentz, A.~W.~Thomas,
  J.\ Phys.\ G {\bf G32}, 667 (2006).

\bibitem{Bentz:2002um}
  W.~Bentz, T.~Horikawa, N.~Ishii, A.~W.~Thomas,
  Nucl.\ Phys.\  {\bf A720}, 95 (2003).

\bibitem{Bentz:2001vc}
  W.~Bentz, A.~W.~Thomas,
  Nucl.\ Phys.\  {\bf A696}, 138-172 (2001).

\bibitem{Bazavov:2011nk} 
  A.~Bazavov, 
 {\it et al.},
  Phys.\ Rev.\ D {\bf 85}, 054503 (2012).

\bibitem{Borsanyi:2010bp} 
  S.~Borsanyi {\it et al.}  [Wuppertal-Budapest Collaboration],
  JHEP {\bf 1009}, 073 (2010).

\bibitem{Fukushima:2003fw} 
  K.~Fukushima,
  Phys.\ Lett.\ B {\bf 591}, 277 (2004).

\bibitem{Ratti:2005jh} 
  C.~Ratti, M.~A.~Thaler and W.~Weise,
  Phys.\ Rev.\ D {\bf 73}, 014019 (2006).

\bibitem{Megias:2004hj} 
  E.~Megias, E.~Ruiz Arriola and L.~L.~Salcedo,
  Phys.\ Rev.\ D {\bf 74}, 065005 (2006).

\bibitem{Hansen:2006ee} 
  H.~Hansen, W.~M.~Alberico, A.~Beraudo, A.~Molinari, M.~Nardi and C.~Ratti,
  Phys.\ Rev.\ D {\bf 75}, 065004 (2007).

\bibitem{Buballa:2003qv}
  M.~Buballa,
  Phys.\ Rept.\  {\bf 407}, 205 (2005).

\bibitem{Klevansky:1992qe}
  S.~P.~Klevansky,
  Rev.\ Mod.\ Phys.\  {\bf 64}, 649 (1992).

\bibitem{Hatsuda:1994pi} 
  T.~Hatsuda and T.~Kunihiro,
  Phys.\ Rept.\  {\bf 247}, 221 (1994).

\bibitem{Vogl:1991qt} 
  U.~Vogl and W.~Weise,
  Prog.\ Part.\ Nucl.\ Phys.\  {\bf 27}, 195 (1991).
 
\bibitem{Kapusta}
J.~I.~Kapusta, {\it Finite Temperature Field Theory}
(Cambridge University Press, Cambridge, 1989).

\bibitem{Klahn:2006iw} 
  T.~Kl\"ahn, D.~Blaschke, F.~Sandin, C.~Fuchs, A.~Faessler, H.~Grigorian, 
 G.~R\"opke and J.~Tr\"umper,
  Phys.\ Lett.\ B {\bf 654}, 170 (2007).

\bibitem{Ebert:2004dr} 
  D.~Ebert, K.~G.~Klimenko and V.~L.~Yudichev,
  Phys.\ Rev.\ C {\bf 72}, 015201 (2005).
 
\bibitem{Blaschke:2004cs} 
  D.~Blaschke, D.~Ebert, K.~G.~Klimenko, M.~K.~Volkov and V.~L.~Yudichev,
  Phys.\ Rev.\ D {\bf 70}, 014006 (2004).

\bibitem{Nielsen:1975hm} 
  H.~B.~Nielsen and S.~Chadha,
  Nucl.\ Phys.\ B {\bf 105}, 445 (1976).

\bibitem{Watanabe:2011ec} 
  H.~Watanabe and T.~Brauner,
  Phys.\ Rev.\ D {\bf 84}, 125013 (2011).

\bibitem{Brauner:2007uw} 
  T.~Brauner,
  Phys.\ Rev.\ D {\bf 75}, 105014 (2007).

\bibitem{Dietrich:2003nu} 
  D.~D.~Dietrich and D.~H.~Rischke,
  Prog.\ Part.\ Nucl.\ Phys.\  {\bf 53}, 305 (2004).

\bibitem{Zablocki:2012zz} 
  D.~S.~Zablocki, D.~Blaschke and M.~Buballa,
  Phys.\ Atom.\ Nucl.\  {\bf 75}, 910 (2012).

\bibitem{Strodthoff:2011tz} 
  N.~Strodthoff, B.~-J.~Schaefer and L.~von Smekal,
  Phys.\ Rev.\ D {\bf 85}, 074007 (2012).

\bibitem{Kleinhaus:2007ve} 
  V.~Kleinhaus, M.~Buballa, D.~Nickel and M.~Oertel,
  Phys.\ Rev.\ D {\bf 76}, 074024 (2007).

\bibitem{Beth:1937zz}
  E.~Beth, G.~Uhlenbeck,
  Physica {\bf 4}, 915 (1937).

\bibitem{Beth:1936zz}
  G.~Uhlenbeck, E.~Beth, 
  Physica {\bf 3}, 729 (1936).
  

\bibitem{Zimmermann:1985ji}
  R.~Zimmermann, H.~Stolz,
  physica\ status\ solidi\ (b) {\bf 131}, 151 (1985).


\bibitem{KKER}  
W.~D.~Kraeft, D.~Kremp, W.~Ebeling, and G.~R\"opke, 
{\it Quantum Statistics of Charged Particle Systems} 
(Plenum, New York, and Akademie-Verlag, Berlin, 1986). 

\bibitem{RMS}
G.~R\"opke, L.~M\"unchow, H.~Schulz, 
Nucl. Phys. {\bf A 379 }, 536 (1982).  

\bibitem{Levinson}
N.~Levinson,
D. Kgl. Danske Vidensk. Selskab. Mat.-fys. Medd. XXXV, Nr. 9 (1949).

\bibitem{Dashen:1969ep} 
  R.~Dashen, S.~-K.~Ma and H.~J.~Bernstein,
  Phys.\ Rev.\  {\bf 187}, 345 (1969).

\bibitem{Schmidt:1990}
  M.~Schmidt, G.~R\"opke, H.~Schulz,
  Annals Phys.\  {\bf 202}, 57 (1990).

\bibitem{Abuki:2006dv}
  H.~Abuki,
  Nucl.\ Phys.\  {\bf A791}, 117 (2007).

\bibitem{Zhuang:1994dw} 
  P.~Zhuang, J.~H\"ufner and S.~P.~Klevansky,
  Nucl.\ Phys.\ A {\bf 576}, 525 (1994).

\bibitem{Ropke:2012qv} 
G.~R\"opke, N.~-U.~Bastian, D.~Blaschke, T.~Kl\"ahn, S.~Typel and H.~H.~Wolter,
  Nucl.\ Phys.\ A {\bf 897}, 70 (2013).

\end{thebibliography}
\end{document}